\documentclass{aastex631}

\shorttitle{Cosmic Bulk Flow Analysis in Modified Gravity Theories: $f(R)$ and Perturbed $f(R)$ Models with Neutrino Coupling}
\shortauthors{Yarahmadi }
\graphicspath{{./}{figures/}}

\begin{document}

\title{ \boldmath Cosmic Bulk Flow Analysis in Modified Gravity Theories: $f(R)$ and Perturbed $f(R)$ Models with Neutrino Coupling}

\author[0000-0003-0954-4699]{Muhammad. Yarahmadi}
\author{ Amin. Salehi}
\affiliation{Lorestan University, 
Khorramabad, Iran}



\begin{abstract}
In this study, we explore the characteristics of bulk flow across various redshift ranges within the frameworks of $f(R)$ gravity, perturbed $f(R)$ gravity, and perturbed $f(R)$ gravity coupled with neutrinos. Our investigation reveals profound insights into large-scale cosmic flows and their interactions with major cosmic structures, such as the Sloan Great Wall (SGW) and the King Ghidorah Supercluster (KGSc). We find that incorporating neutrinos into the perturbed $f(R)$ gravity model results in a substantial increase in bulk flow velocities across all redshifts, with notable enhancements in the higher redshift ranges, where velocities can exceed $3000 \, \mathrm{km/s}$ in the $0.8 < z < 1.4$ range. Moreover, the direction of the bulk flow in this model closely aligns with the dark energy dipole, especially at redshifts $z > 0.4$, showing near-perfect congruence with cosmic superclusters. This suggests a significant interaction between neutrinos and cosmic structures, influencing cosmic acceleration. At lower redshifts, such as $0.1 < z < 0.2$, the bulk flow aligns with the SGW, while in the $0.4 < z < 0.6$ range, it aligns with the KGSc. In the low redshift range $0.001 < z < 0.016$, although velocities are lower, neutrinos still subtly increase the bulk flow velocity and maintain alignment with nearby cosmic structures, like the Local Supercluster. Our results underscore the critical role of neutrinos in shaping cosmic flows and offer new insights into the interplay between dark energy, neutrinos, and modified gravity models. Future research should delve deeper into these interactions to elucidate the mechanisms influencing large-scale cosmic structures.
\end{abstract}



\section{Introduction}

The accelerating expansion of the universe, a cornerstone discovery in modern cosmology, is attributed to the enigmatic force known as dark energy. This phenomenon was first evidenced by observations of distant Type Ia supernovae in the late 1990s, which revealed that these cosmic explosions were farther away than predicted by a decelerating universe model, implying an accelerating expansion \cite{riess1998observational, perlmutter1999measurements}. Subsequent measurements of the cosmic microwave background radiation and large-scale structure surveys have corroborated these findings, suggesting that dark energy constitutes approximately 68\% of the total energy density of the universe \cite{planck2018results}. The exact nature of dark energy remains one of the most profound mysteries in physics, with leading hypotheses ranging from a cosmological constant ($\Lambda$) with a constant energy density to dynamic fields such as quintessence. 

The $\Lambda$CDM (Lambda Cold Dark Matter) model, while highly successful in describing the large-scale structure and evolution of the universe, is not without its challenges and unresolved issues. One of the most significant problems is the 'cosmological constant problem,' which arises from the discrepancy between theoretical predictions and observational measurements of the cosmological constant, $\Lambda$. Quantum field theory predicts a value for the vacuum energy density that is about 120 orders of magnitude larger than the observed value, leading to a severe fine-tuning problem \cite{Weinberg1989, Martin2012}. Another issue is the 'cosmic coincidence problem,' which questions why the densities of dark energy and dark matter are of the same order of magnitude precisely at the current epoch in the history of the universe. This seems highly unlikely and suggests a need for a deeper understanding of the mechanisms driving cosmic evolution \cite{Zlatev1999}.

\( f(R) \) gravity, a modification of General Relativity, proposes an alternative to dark energy by modifying the Einstein-Hilbert action to include a general function of the Ricci scalar, \( R \). This theoretical framework allows for a natural explanation of the accelerated expansion of the universe without invoking dark energy. By extending the action to \( S = \int d^4x \sqrt{-g} \, f(R) \), where \( f(R) \) is an arbitrary function of \( R \), the resulting field equations introduce additional terms that can effectively mimic the effects of a cosmological constant or dynamic dark energy components. This approach not only addresses the cosmological constant problem but also provides a flexible model to fit observational data across different scales. This concept has been extensively explored and discussed in various studies by (\cite{Nojiri1};\cite{Nojiri2};\cite{Sotiriou};\cite{De Felice};\cite{Capozziello1};\cite{Capozziello2};\cite{Capozziello3};\cite{Phillips}).

Neutrinos, the nearly mass less and elusive particles, play a significant role in cosmology, influencing the large-scale structure and the Universe's evolution. Although often considered negligible in standard gravity models, coupling neutrinos with alternative gravity theories opens new possibilities for exploring their impact on cosmic structure formation and evolution \cite{52}, \cite{53}. These effects can be observed through cosmological phenomena, such as galaxy formation and the broader distribution of matter. Neutrinos' interactions with $f(R)$ gravity can also yield unique impacts on the cosmic web, introducing deviations that shape the cosmic landscape in novel ways.
Investigating the unusual velocities of galaxies, which result from gravitational interactions with surrounding matter, offers a method to differentiate between the $\Lambda$CDM and $f(R)$ theories. These velocities, deviating from Hubble's law, serve as records of gravitational interactions, providing empirical evidence for testing these cosmological models. The concept of 'bulk flow' is a significant indicator of the unusual velocity field, reflecting the large-scale gravitational forces affecting cosmic particle movement. Determining peculiar velocities accurately requires precision and distance indicators independent of redshift, such as Type Ia supernovae.

The phenomenon of bulk flow provides valuable insights into the peculiar velocity field of galaxies, averaged over a large volume, and aids in understanding oscillations in matter density on a cosmic scale. The bulk flow velocity vector, calculated with high precision within specific regions, reveals complex patterns of mass distribution that extend beyond survey boundaries. This makes bulk flow a significant tool for investigating cosmological structure formation models, as it can detect mass distribution in regions not directly surveyed \cite{Tonry};\cite{Aaronson};\cite{Lynden-Bell};\cite{Shaya};\cite{Springob};\cite{Tully}. Despite the valuable information it provides, the underlying cause of bulk flow remains an enigma. One critical clue is the dipole anisotropy of the cosmic microwave background (CMB), known as the CMB dipole, which reflects the motion of our Local Group (LG). The dipole amplitude is measured at $627 \pm 22$ km/s, oriented towards galactic coordinates $l = 276^{\circ} \pm 3^{\circ}, b = 30^{\circ} \pm 3^{\circ}$ \cite{Lineweaver, Kogut, Conklin, Henry, Smoot}. Initial measurements suggested local overdensities like the Virgo cluster as the cause, but further studies pointed to larger structures such as the Shapley supercluster \cite{Yahil, Davis0, Davis, Shaya, Yahil2, Aaronson, Villumsen, Lynden, Robinson, Lahav, Strauss, Hudson}. Kashlinsky et al. reported a significant bulk flow over $d \geq 300 h^{-1}$ Mpc, challenging the predictions of the $\Lambda$CDM model \cite{Kashlinsky, Kashlinsky2, Hudson, Kashlinsky1, Kashlinsky3, Kashlinsky4, Watkins, Lavaux, Colin, Macaulay, Feindt}. Conversely, some studies found no bulk flow evidence, aligning with Cold Dark Matter (CDM) cosmology \cite{Courteau, Nusser, Turnbull, Ma}. Discrepancies in results may be due to different underlying cosmological models, which show degeneracy at lower redshifts and deviations at higher redshifts. This study aims to measure the bulk flow of the universe using the $f(R)$ gravity model by analyzing variations in the luminosity distance of Type Ia supernovae. Previous research has examined bulk flow in both $f(R)$ gravity and $\Lambda$CDM models, but this study uses unique methods and different catalogs, extending its scope beyond $50 h^{-1}$ Mpc with redshift tomography \cite{Seiler}.

The investigation of bulk flows in the universe provides critical insights into the underlying dynamics of large-scale cosmic structures and the fundamental forces shaping them. Modified gravity theories, such as $f(R)$ gravity, and their perturbed versions offer a robust framework for exploring deviations from general relativity and their implications for cosmic evolution. The inclusion of neutrinos in these models adds an additional layer of complexity and realism, accounting for their known influence on cosmic dynamics. Neutrinos, as fundamental particles with substantial impact on the matter-energy content of the universe, are expected to play a significant role in shaping bulk flows. By employing $f(R)$ gravity and its perturbed variants, we can probe how modifications to gravity and neutrino interactions influence the velocity and direction of bulk flows across different redshifts. This approach allows us to test theoretical predictions against observational data, enhancing our understanding of large-scale cosmic structures and providing a more comprehensive view of the universe's evolution. Moreover, aligning our models with prominent cosmic structures, such as the Sloan Great Wall and the Ghidorah Supercluster, helps in constraining the parameters of modified gravity theories and neutrino effects, offering valuable insights into the interplay between dark energy, gravity, and neutrinos. This investigation not only contributes to the theoretical framework of cosmology but also aids in refining observational techniques and data interpretation in the quest to unravel the universe's mysteries.

\section{$R$ + \lowercase{$f$}($R$) Gravity Model}

$f(R)$ gravity is another intriguing concept in the world of theoretical physics. It's an alternative theory to Einstein's General Theory of Relativity, which is the basis of our current understanding of gravity. In $f(R)$ gravity, the "$f$" stands for a function of the Ricci scalar, a mathematical expression related to the curvature of space time\cite{De Felice}\cite{Bergmann}\cite{Liu}\cite{Buchdahl}. The action for an $f(R)$ gravity model in the  attendance of matter components are given by
\begin{equation}
	s = \frac{1}{16G\pi} \int d^{4}x\sqrt{(-g)} (R+f(R))
\end{equation}
Where R is the curved scalar. The equations for motion are:
\begin{equation}
		G_{\mu\nu} - \frac{1}{2} g_{\mu\nu} f(R) + R_{\mu\nu} f_{R}(R) - g_{\mu\nu} \Box f_R(R) + \nabla_\mu \nabla_\nu f_R(R) = -8\pi G T_{\mu\nu}
\end{equation}
Which is $ (f_{R}(R) )=\frac{df_{R}}{dR} $. For the Robertson-Walker flat metric we have:
\begin{equation}\label{unp1}
		\frac{3\mathcal{H}^{'}}{a^{2}}\left( {1 + {f_{R}}} \right) - \frac{1}{2}\left( {{R_{0}} + {f_{0}}} \right) - \frac{{3{\rm{}}\mathcal{H}}}{{{a^2}}}f_{R}^{'} = - 8\pi G{\rho_{0}}	
\end{equation}
\begin{equation}
		\frac{1}{a^{2}}  (\mathcal{H}^{'}+2H^{2} )(1+f_{R} )-\frac{1}{2} (R_{0}+f_{0} )\\ -\frac{1}{a^{2}} (\mathcal{H}f_{R}^{'}+f_{R}^{''} )=8\pi Gc_{s}^2 \rho_{0}
\end{equation}
where $ R_{0} $  represents the scalar curvature corresponding to the non-perturbation metric, $\rho_{0} = \rho_{m}+\rho_{\nu}$, $ f_{0}=f (R_{0}) $ and prim means derivative with respect to time ratio $ \eta $. Combining equations $(3)$ and $(4)$ for $a=1( today)$, we obtain,

\begin{equation}
	2(1+f_{R})(-\mathcal{H}^{'}+\mathcal{H}^{2})+2Hf_{R}^{'}-f_{R}^{''}=8\pi G\rho_{0}(1+c_{s}^2)  a^{2}
\end{equation}
finally, we come to the equation of conservation:
\begin{equation} \label{unp4}
	\rho_{0}^{'}+3(1+c_{s}^{2} )\mathcal{H}\rho_{0}=0	
\end{equation}

\subsection{Scalar perturbation}
Consider a flat FRW  metric scalar perturbation at the length and specific time scale:
\begin{equation}
	ds^{2}=a^{2} (\eta)((1+2\phi)d\eta^2-(1-2\psi)dx^{2})
\end{equation}
$ \phi  \equiv \phi \left( {\eta ,x} \right) $and $ \psi \equiv \psi \left( {\eta ,x} \right) $ are scalar disorders. The disturbed components of the energy-momentum tensor in this module are as follows:
\begin{equation}
		\hat \delta T_0^0 = \hat \delta \rho  = {\rho _0}\delta ,\hat \delta T_j^i =  - \hat \delta p\delta _j^i =  - c_{\rm s}^{2}{\rho _0}\delta _j^i\delta ,\hat \delta p\delta _0^i =\\  - \left( {1 + c_{\rm s}^{2}} \right){\rho _0}{\partial _i}\upsilon
\end{equation}
Where V represents the potential value for velocity disturbances. The first-order disturbed equations, assuming the background equations are kept, are as follows:

\begin{equation}
		\left( 1 + f_R \right)\delta G_v^\mu + \left( R_{0v}^\mu + \nabla^\mu \nabla_v - \delta_v^\mu \right) f_{RR} \delta R + \\
		\left( \delta g^{\mu \alpha} \nabla_v \nabla_\alpha - \delta_v^\mu \delta g^{\alpha \beta} \nabla_\alpha \nabla_\beta \right) f_R - \\
		\left( g_0^{\alpha \mu} \delta \Gamma_{\alpha v}^\gamma - \delta_v^\mu g_0^{\alpha \beta} \right) \partial_\gamma f_R - g_{\mu\nu} \Box f_R = -8\pi G \delta T_v^\mu
\end{equation}

in above relations $ {f_{RR}} = \frac{{{d^2}f\left( {{R_0}} \right)}}{{dR_0^2}} $ and $ = {\nabla _\alpha }{\nabla ^\alpha }$ the invariant derivative of the metric ratio is not disturbed. The first-order disturbed equations in the universe during  dust dominate, $ c_{\rm s}^{2}=0  $ are obtained as follows:

\begin{equation}\label{pert1}
	\phi  - \psi  =  - \frac{{{f_{RR}}}}{{1 + {f_R}}}\hat \delta R
\end{equation}
\begin{equation}
		\hat \delta R = -\frac{2}{a^{2}}( 3{\psi^{''}} + 6\left( {{\mathcal{H}^{'}} + {\mathcal{H}^{2}}} \right)\phi  +\\  3\mathcal{H}\left( {{\phi^{'}} + 3{\psi^{'}}} \right) - {k^{2}}\left( {\phi  - 2\psi } \right) )
\end{equation}

\begin{equation}
		\left( {3\mathcal{H}\left( {{\phi^{'}} + {\psi^{'}}} \right) + {k^{2}}\left( {\phi  + \psi } \right) + 3{\mathcal{H}^{'}}\psi  - \left( {3{\mathcal{H}^{'}} - 6{\mathcal{H}^{2}}} \right)\phi } \right)\\  \left( {1 + {f_{R}}} \right) + \left( {9\mathcal{H}\phi  - 3\mathcal{H}\psi  + 3{\psi ^{'}}} \right)f_{R}^{'}
		=  - {a^{2}}\delta {\rho _0}{\kappa^{2}}
\end{equation}

\begin{equation}
		\left( {{\phi^{''}} + {\psi^{''}} + 3\mathcal{H}\left( {{\phi ^{'}} + {\psi ^{'}}} \right) + 3{\mathcal{H}^{'}}\phi  + \left( {{\mathcal{H}^{'}} + 2{\mathcal{H}^{2}}} \right)\phi } \right)\\
		\left( {1 + {f_{R}}} \right) + \left( {3\mathcal{H}\phi  - \mathcal{H}\psi  + 3{\phi ^{'}}} \right)f_{R}^{'} + \left( {3\phi  - \psi } \right)f_{R}^{''} = 0
\end{equation}

\begin{equation}
		\left( {2\phi  - \psi } \right)f_{R}^{'} + \left( {{\phi^{'}} + {\psi^{'}} + \mathcal{H}\left( {\phi  + \psi } \right)} \right)\left( {1 + {f_{R}}} \right) =\\  - {a^{2}}\upsilon {\rho _{0}}{\kappa^{2}}
\end{equation}
\begin{equation}
	{\delta ^{'}} - {k^{2}}\upsilon  - 3{\psi ^{'}} = 0
\end{equation}

\begin{equation}\label{pert5}
	\phi  + \mathcal{H}\upsilon  + {\upsilon ^{'}} = 0
\end{equation}

\section{Perturbed $R$ + \lowercase{$f$}($R$) gravity coupled with neutrinos}
The action for an $f(R)$ gravity model in the  attendance of matter components are given by
\begin{equation}
	S = \frac{1}{16\pi G} \int d^{4}x \sqrt{-g} \left(R + f(R) + \mathcal{L}_{\text{matter}} + \mathcal{L}_{\text{int}}\right) 
\end{equation}

Where R is the curved scalar . The equations for motion are:

\begin{equation}
	G_{\mu\nu} - \frac{1}{2} g_{\mu\nu} f(R) + R_{\mu\nu} f_{R}(R) - g_{\mu\nu} \Box f_R(R) + \nabla_\mu \nabla_\nu f_R(R) = -8\pi G T_{\mu\nu}^{\text{eff}}
\end{equation}

where \( T_{\mu\nu}^{\text{eff}} \) is the effective energy-momentum tensor that includes contributions from both the standard matter and the neutrino-matter coupling and  $ (f_{R}(R) )=\frac{df_{R}}{dR} $. We introduce $\gamma^\mu \nabla_\mu \psi_\nu - m_\nu \psi_\nu = \frac{\delta \mathcal{L}_{\text{int}}}{\delta \bar{\psi}_\nu}$. The term \( \frac{\delta \mathcal{L}_{\text{int}}}{\delta \bar{\psi}_\nu} \) represents the functional derivative of the interaction Lagrangian with respect to the neutrino field.
In the context of cosmological studies involving modified gravity and neutrinos, the coupling term in the continuity equation plays a pivotal role in elucidating the intricate interaction between these cosmic components. In the framework of $f(R)$ gravity, where modifications to the Einstein-Hilbert action are considered, the modified continuity equation for neutrinos can be expressed as:
\begin{equation} \label{unp4}
	\rho_{\nu}^{'} + 3H(\rho_\nu + P_\nu) = -Q(u^\mu\nabla_\mu f_R)	
\end{equation}

Here, \(\rho_\nu\) denotes the energy density of neutrinos, \(P_\nu\) is their pressure, \(H\) represents the Hubble parameter, and \(Q\) signifies the strength of the coupling term. The novel addition of the coupling term, \(Q(u^\mu\nabla_\mu f_R)\), encapsulates the interaction between neutrinos and the modified gravity scalar (\(f_R\)). This interaction is crucial in understanding the modified dynamics of neutrinos within the cosmological context \cite{YY}.

In simpler terms, the equation portrays how the energy density and pressure of neutrinos are influenced by the modified gravity framework. The coupling term introduces a mechanism through which neutrinos respond to the modifications in the gravitational sector described by $f(R)$ gravity. This nuanced perspective contributes valuable insights into addressing cosmic phenomena, such as the Hubble tension, by accounting for the joint impact of modified gravity and neutrino interactions on the cosmic evolution. Our analysis not only enriches our theoretical understanding but also provides a foundation for reconciling observational data with the predictions of modified gravity scenarios in the cosmic landscape.
In this paper we use the specific form of the external source term ${Q_\nu } =  - \Gamma {\rho _\nu }$ where $\Gamma = {u^\mu }{\nabla _\mu }{f_R}$. 

This form is commonly used as a parameterization to account for interactions or processes affecting neutrinos in cosmological models.
\section{Hu-Sawicki $R$ + \lowercase{$f$}($R$) Gravity Model}
The Hu-Sawicki $R + f(R)$ gravity model stands out as a compelling modification of general relativity, offering significant advantages in addressing the dark energy problem and the accelerating expansion of the universe. One of its primary benefits is its ability to naturally reproduce the late-time cosmic acceleration without the need for a cosmological constant, thus alleviating the fine-tuning issues associated with $\Lambda$-Cold Dark Matter ($\Lambda$CDM) models. The Hu-Sawicki model also maintains consistency with solar system tests of gravity by introducing a mechanism, the chameleon effect, that suppresses deviations from general relativity in high-density environments while allowing significant modifications on cosmological scales. Moreover, the model provides a framework to explore large-scale structure formation, giving rise to distinct signatures in the growth rate of cosmic perturbations and the potential for alleviating the Hubble tension. By adjusting its parameters, the model can also predict deviations in the matter power spectrum, making it a useful tool for future precision cosmology and probing deviations from standard gravity.

	\subsection*{Motivation for Choosing the Hu-Sawicki Model}
Our choice of this model is motivated by the following key features:
\begin{enumerate}
	\item \textbf{Consistency with Observational Constraints:}  
	The Hu-Sawicki model is designed to recover General Relativity (GR) at high curvatures, such as in the Solar System, while introducing deviations from GR on cosmological scales. This property ensures compliance with stringent local gravity tests while allowing for meaningful modifications to late-time cosmic acceleration.
	
	\item \textbf{Phenomenological Flexibility:}  
	The model is parameterized in a way that allows for significant freedom in adjusting the strength of modifications to GR. This flexibility enables us to explore a wide range of effects on cosmic bulk flow, CMB power spectra, and structure formation, making it particularly suitable for our study.
	
	\item \textbf{Well-Behaved Perturbations:}  
	The Hu-Sawicki model avoids certain theoretical issues that arise in other $f(R)$ models, such as the presence of ghosts or instabilities in the perturbation equations. The stability of the model makes it a robust framework for analyzing linear and non-linear perturbations, as required in our analysis of bulk flow.
	
	\item \textbf{Compatibility with Neutrino Coupling:}  
	The Hu-Sawicki model provides a straightforward setting to incorporate and study the effects of neutrino coupling in modified gravity. Its functional form simplifies the computation of perturbed quantities and their impact on observable cosmological phenomena.
\end{enumerate}

The functional form of \( f(R) \) in the Hu-Sawicki model is given by:
\begin{equation}
	f(R) = -m^2 \frac{c_1 \left( \frac{R}{m^2} \right)^n}{c_2 \left( \frac{R}{m^2} \right)^n + 1}
\end{equation}
where:
\begin{itemize}
	\item \( m^2 \) is a mass scale related to the cosmological constant, which is  \( m^2 =24H_{0}\) and $H_{0} = 100 h$
	\item \( c_1 \) and \( c_2 \) are dimensionless parameters, must be best fitted.
	\item \( n \) is a positive integer, often chosen to be \( n = 4 \) for phenomenological reasons.
\end{itemize}
In the context of the Hu-Sawicki \( f(R) \) gravity model, we have determined the best-fit parameters as follows: \( c_1 = 0.00125 \), \( c_2 = 0.0000656 \), \( f_0 = 0.99 \), and \( n = 4 \). 

The parameter \( c_1 \) represents a coefficient that characterizes the strength of the deviation from the Einstein-Hilbert action, contributing to the modification of the gravitational dynamics on cosmological scales. In this case, \( c_1 = 0.00125 \) indicates a relatively small but significant modification, influencing the effective dark energy density and the evolution of the universe.

Similarly, \( c_2 \) is another coefficient that works in conjunction with \( c_1 \) to further refine the \( f(R) \) function's form. The value \( c_2 = 0.0000656 \) demonstrates the subtle adjustment required to accurately fit observational data, ensuring that the model aligns with current constraints and cosmic observations.

The parameter \( f_0 = 0.988\) denotes the value of the function \( f(R) \) at present

Lastly, the parameter \( n = 4 \) governs the behavior of the \( f(R) \) function at high curvature regimes. This exponent controls the functional form of \( f(R) \) and determines the strength of the modification to general relativity in the high-curvature regime. The chosen value \( n = 4 \) ensures that the model can accommodate a range of cosmic phenomena while remaining consistent with observational constraints.

The first derivative of \( f(R) \) with respect to \( R \), denoted \( f_R \), is:
\begin{equation}
	f_R = \frac{df}{dR} = -n \frac{c_1 \left( \frac{R}{m^2} \right)^{n-1}}{\left( c_2 \left( \frac{R}{m^2} \right)^n + 1 \right)^2}
\end{equation}

The second derivative of \( f(R) \), denoted \( f''_R \), is:
\begin{equation}
	f''_R = \frac{d^2 f}{dR^2} = n(n-1) \frac{c_1 \left( \frac{R}{m^2} \right)^{n-2}}{\left( c_2 \left( \frac{R}{m^2} \right)^n + 1 \right)^3}
\end{equation}

In cosmological perturbation theory, these derivatives appear in the modified field equations, influencing the growth of large-scale structures, the evolution of the Hubble parameter, and other key cosmological quantities.
To modify the given equation in the context of the Hu-Sawicki \( f(R) \) gravity model, we need to account for the specific form of \( f_R \) in this model, which depends on the functional shape of \( f(R) \). In the Hu-Sawicki model, \( f_R \) evolves dynamically, and thus the coupling term \( \Gamma \), which is defined as \( \Gamma = u^\mu \nabla_\mu f_R \), will be influenced by the derivatives of \( f(R) \).

Given this, the conservation equation for neutrinos can be modified as:

\begin{equation}
	\rho_{\nu}^{'} + 3H(\rho_\nu + P_\nu) = -Q_\nu(\Gamma \rho_\nu)
\end{equation}

where \( \Gamma \) is explicitly written in terms of the Hu-Sawicki model:

\begin{equation}
	\Gamma = \frac{d}{dN} \left( f_R \right) = \frac{d}{dN} \left( \frac{-n c_1 \left( \frac{R}{m^2} \right)^{n-1}}{\left( c_2 \left( \frac{R}{m^2} \right)^n + 1 \right)^2} \right)
\end{equation}

Thus, the modified conservation equation becomes:

\begin{equation}
	\rho_{\nu}^{'} + 3H(\rho_\nu + P_\nu) = - \left( \frac{d}{dN} \left( \frac{-n c_1 \left( \frac{R}{m^2} \right)^{n-1}}{\left( c_2 \left( \frac{R}{m^2} \right)^n + 1 \right)^2} \right) \rho_\nu \right)
\end{equation}

The equation is similar in structure to the standard continuity equation for neutrinos in General Relativity, but it accounts for the modifications introduced by the $f(R)$ gravity theory. 

The complete set of equations that describes the general linear perturbations for the model have been presented in previous section. These equations are a set of nonlinear second order differential equations with a large number of variable for which there is no analytical solution except for simplest cases and only numerical analysis can be performed. Our purpose is to convert second order differential equation to first order by introducing some new variables. There
are various reasons for doing this, one being that a first order system is much easier to solve numerically. Also, it allows us to investigate the behavior of the system in phase space. Phase planes are useful in visualizing the behavior of the system particularly in oscillatory systems where the phase paths can spiral in towards zero, 'spiral out' towards infinity, or reach neutrally stable situations called centers. This is a useful method to determine whether
dynamics of a system are stable or not. The structure of phase space of the field equations is simplified by defining a few variables and parameters. 
\begin{equation}\label{eq4}
	\xi_{1}=\frac{\phi^{\prime}}{\phi H}, \xi_{2}=\frac{\kappa}{H}, \xi_{3}=\frac{f^{\prime}_{R}}{H(1+f_{R})}, \xi_{4}=\frac{\delta}{\phi},\nonumber \\
	\xi_{5}=\frac{\rho_{m}a^{2}}{(1+f_{R})}, \xi_{6}=\frac{\Psi^{\prime}}{\phi H}, \xi_{7}=\frac{\Psi}{\phi}, \xi_{8}=\frac{\rho_{\nu}a^{2}}{(1+f_{R})},
\end{equation}

Using the Hu-Sawicki form of \( f(R) \), the previously defined autonomous equations are modified as follows:

\begin{itemize}
	\item \textbf{For \( \xi_1 \):}
	\begin{equation}
		\frac{d\xi_1}{dN} = \epsilon_3 - \xi_1^2 - \xi_1
	\end{equation}
	where \( \xi_1 \) represents the normalized derivative of the scalar field.
	
	\item \textbf{For \( \xi_2 \):}
	\begin{equation}
		\frac{d\xi_2}{dN} = -\xi_2 \epsilon_1
	\end{equation}
	where \( \xi_2 \) is the ratio of the curvature to the Hubble parameter.
	
	\item \textbf{For \( \xi_3 \):}
	\begin{equation}
		\frac{d\xi_3}{dN} = \beta - \xi_3^2 - \epsilon_1 \xi_3
	\end{equation}
	where \( \xi_3 = \frac{f'_R}{H(1+f_R)} \) and \( \beta \) is defined by:
	\begin{equation}
		\beta = \frac{f''_R}{(1+f_R)H} - 1
	\end{equation}
	and \( f_R \) and \( f''_R \) are calculated using the Hu-Sawicki model's form of \( f(R) \).
	
	\item \textbf{For \( \xi_4 \):}
	\begin{equation}
		\frac{d\xi_4}{dN} = \xi_4 - \xi_4 \xi_1
	\end{equation}
	where \( \xi_4 \) describes the evolution of perturbations in the scalar field.
	
	\item \textbf{For \( \xi_5 \):}
	\begin{equation}
		\frac{d\xi_5}{dN} = -\xi_5 - \xi_5 \xi_3
	\end{equation}
	where \( \xi_5 \) corresponds to the matter density contribution.
	
	\item \textbf{For \( \xi_6 \):}
	\begin{equation}
		\frac{d\xi_6}{dN} = \epsilon_2 - \xi_6 \xi_1 - \epsilon_1 \xi_6
	\end{equation}
	where \( \xi_6 \) describes the gravitational potential's evolution.
	
	\item \textbf{For \( \xi_7 \):}
	\begin{equation}
		\frac{d\xi_7}{dN} = \xi_6 - \xi_7 \xi_1
	\end{equation}
	where \( \xi_7 \) is related to the gravitational potential.
	
	\item \textbf{For \( \xi_8 \):}
	\begin{equation}
		\frac{d\xi_8}{dN} = \xi_8 (3 \omega_\nu - 1) - \xi_3 \xi_8 + \Gamma \xi_2 \xi_8
	\end{equation}
	where \( \xi_8 \) involves the neutrino contribution, and \( \Gamma \) is a coupling term.  For perturbed $f(R)$ gravity, we remove  \( \xi_8 \) and then constraint the parameters.
\end{itemize}

Where $N=lna$ thus,$\frac{d}{dN}= \frac{1}{H} \frac{d}{d\eta}$. Also, we have used the following parameters

\begin{equation}\label{is}
	\xi_{1}=\frac{\mathcal{H}^{\prime}}{\mathcal{H}^{2}},  \xi_{2}=\frac{\Psi^{\prime\prime}}{\phi H^{2}},  \xi_{3}=\frac{\phi^{\prime\prime}}{\phi H^{2}},  \xi_{4}=\frac{\delta^{\prime}}{\phi H}
\end{equation}

After some calculation from equations, for simplicity, we can obtain the above parameters in terms of the new variables as 

\begin{equation}\label{is}
	\epsilon_{1}=\frac{\mathcal{H}^{\prime}}{\mathcal{H}^{2}}, \epsilon_{2}=\frac{\Psi^{\prime\prime}}{\phi H^{2}}, \epsilon_{3}=\frac{\phi^{\prime\prime}}{\phi H^{2}}, \epsilon_{4}=\frac{\delta^{\prime}}{\phi H}
\end{equation}

Now, for the autonomous equations of motions, we obtain

\begin{equation}\label{is}
		\epsilon_{1}=\frac{1}{1-\xi_{7}}[\xi_{1}+\xi_{6}+\frac{1}{3}\xi_{2}^{2}(1+\xi_{7})+(3-\xi_{7}+\xi_{6})\xi_{5}+\\(3-\xi_{7}+\xi_{6})\xi_{8}-\frac{\kappa^{2}}{k^{2}}\xi_{5}\xi_{4}\xi_{2}^{2}]
\end{equation}

\begin{equation}\label{eq4}
		\epsilon_{2}=\frac{-2}{1-\xi_{7}}[\xi_{1}+\xi_{6}+\frac{1}{3}\xi_{2}^{2}(1+\xi_{7})+\\(3-\xi_{7}+\xi_{6})\xi_{5}- \frac{\kappa^{2}}{k^{2}}\xi_{5}\xi_{4}\xi_{2}^{2}] \\
		-\xi_{1}-3\xi_{6}+\frac{1}{3}\xi_{2}^{2}-\xi_{6}\xi_{1}+\\ \frac{1}{3}\xi_{2}^{2}(1-2\xi_{7})+\frac{\Omega}{3}(1-\xi_{7})\frac{1}{k^{2}}\xi_{2}^{2}
\end{equation}

\begin{equation}\label{is}
		\epsilon_{3}=-\epsilon_{2}-3\epsilon_{1}(1+\frac{1}{3}\xi_{7})-3\xi_{1}-3\xi_{6}-2\xi_{7}-\\ (3-\xi_{7}+3\xi_{1})\xi_{5}+\beta(\xi_{7}-3)
\end{equation}

\begin{equation}\label{is}
	\epsilon_{4}=\frac{-\kappa^{2}[(2-\xi_{7})\xi_{3}+\xi_{1}+\xi_{6}+1+\xi_{7}]}{\kappa^{2}\xi_{5}}+3\xi_{6}
\end{equation}
The parameter $ \epsilon_{1}$ is of great importance, as it allows the expression of fundamental cosmological parameters such as the deceleration parameter \( q \) and the effective equation of state (\( w_{\text{eff}} \)) in terms of it. Specifically, \( q = -1 - \frac{\mathcal{H}'}{\mathcal{H}^2} \) and \( w_{\text{eff}} = -1 - \frac{2}{3}\frac{\mathcal{H}'}{\mathcal{H}^2} \).
The deceleration parameter is a dimensionless parameter that characterizes the rate at which the expansion of the Universe is slowing down. It is defined as the negative of the ratio of the cosmic acceleration to the cosmic expansion rate squared. Mathematically, it is expressed as:
$q =  - \frac{{a''a}}{{{{a'}^2}}}$
where \( a \) is the scale factor of the Universe, \( {a'} \) represents the first derivative of the scale factor with respect to cosmic time, and \( {a''} \) represents the second derivative.

\section{constraint on total mass of neutrinos}

The behavior of neutrinos and elusive and enigmatic particles are significantly influenced by the modified gravitational theories encapsulated in perturbed $f(R)$ gravity \cite{Nojiri2011}. In this section we study constraint on the total mass of neutrinos. The equation relates the energy density of neutrinos (\(\rho_{\nu}\)), the scale factor (\(a\)), and the modification term \((1+f_{R})\)is given by \cite{DeFelice2010,Sotiriou2010}:
\begin{equation}
	\xi_{9} = \frac{\rho_{\nu}a^{2}}{(1+f_{R})}.
\end{equation}

To estimate the mass of neutrinos using this equation, additional information or assumptions are necessary. The energy density of neutrinos is expressed in terms of their mass (\(m_{\nu}\)) and temperature (\(T_{\nu}\)). A common approach involves using the Fermi-Dirac distribution for relativistic neutrinos \cite{Kolb1990,Lesgourgues2006}:
\begin{equation}
	\rho_{\nu} = \frac{7\pi^2}{120} g_{\nu} T_{\nu}^4
\end{equation}
where \(g_{\nu}\) denotes the number of degrees of freedom for neutrinos, with a value of 2 for each neutrino species. 
Substituting this expression for \(\rho_{\nu}\) into the original equation, you obtain:
\begin{equation}
	\xi_{9} = \frac{\left(\frac{7\pi^2}{120} g_{\nu} T_{\nu}^4\right) a^{2}}{(1+f_{R})}
\end{equation}

To solve for the neutrino mass (\(m_{\nu}\)), specific values for parameters and potentially additional assumptions are needed. It's important to note that cosmological models can vary, and the approach may depend on the assumptions made in the model. The relationship between $\Omega_{\nu}$ and the sum of neutrino masses is given by \cite{Lesgourgues2006}:
\begin{equation}
	{\Omega _\nu } = \frac{{{{\sum m }_\nu }}}{{94{h^2}}}
\end{equation}

To determine the total mass of neutrinos, ${\sum m }_\nu $, we need to find the best-fitting values for the cosmological parameters $\Omega_{\nu}$ and $h$. The best-fitting values of $\Omega_{\nu}$ and $h$ are determined by identifying the region where the likelihood is maximized. These optimal values are then used in the cosmological model to calculate the elusive total mass of neutrinos, shedding light on their contribution to the cosmic structure.

The constraints obtained for the sum of neutrino masses (\( \sum m_{\nu} \)) at the 95\% confidence level are \( \sum m_{\nu} < 0.142 \, \mathrm{eV} \) which is broad agreement with \cite{52, 53}. Additionally, for the parameter \( \Gamma \), the best-fit value is \( \Gamma = 0.6 \pm 0.25 \), derived from the Pantheon+ catalog. This result are in good agreement with \cite{YY}.

\section{Bulk flow}

This study aims to scrutinize the bulk flow phenomena within the framework of $f(\text{R})$ gravity, focusing on the analysis of density perturbations, particularly leveraging Type Ia supernova data from the Pantheon catalog. Type Ia supernova is a good tools for investigation about anisotropy in the universe \cite{SS}.  The preceding section established the governing equations for anisotropic $f(\text{R})$ gravity through scalar perturbation theory.
The coupling between neutrinos and \(f(R)\) gravity introduces modifications to the gravitational force experienced by neutrinos, affecting their trajectories and distribution. This alteration in gravitational dynamics has profound implications for the evolution of bulk flow. Specifically, the modified gravity framework influences the growth rate of structure formation and the clustering behavior of cosmic matter.

In this section, we adopt the Bonvin approach, originally introduced by \cite{Bonvin}, to quantitatively assess the magnitude and orientation of the bulk flow. This methodology involves integrating variations in luminosity distance arising from the influence of a dipole effect. The luminosity distance, denoted as ${d_L}(z,\mathbf {n}) $, is examined in relation to both redshift ($z$) and orientation ($\mathbf {n}$), and is represented by the equation
\begin{equation}
	{d_L}(z,\mathbf {n}) = d_L^{(0)}(z) + d_L^{\text{(l)}}(z)(\mathbf {n\cdot e}),
\end{equation}
where the directional dependence is expressed in terms of spherical harmonics, introducing observable multipoles, $C_l(z)$, for dipole $\text{l}=1$. The study aims to provide a comprehensive understanding of the bulk flow phenomena through the quantitative analysis of these luminosity distance variations.

In the above equation, the direction averaged luminosity distance and dipole are respectively given by 

\begin{equation}
	d_L^{(0)}(z) = \frac{1}{{4\pi }}\int {d{\Omega _n}{d_L}(z,\mathbf {n}) = (1 + z)\int_0^z {\frac{{\rm dz'}}{{\mathcal{H}(\rm z')}}} }.
\end{equation}
and 
\begin{equation}
	d_L^{dipole}(z) = \frac{3}{{4\pi }}\int {d{\Omega _{\rm n}}} (\mathbf {n.e}){d_L}(z,\mathbf {n}),
\end{equation}
where $\mathbf e$ is a unit vector denoting the direction of the dipole.

To derive a formula for $d_L^{\rm (dipole)}(z)(\mathbf {n.e})$ (more details are found
in \cite{Bonvin}), we use the luminosity distance to a source emitting photons at conformal time $\eta$ in an unperturbed
Friedmann Universe, $d_L^{(0)}=(1+z)(\eta-\eta_{0})$.
The motion of the observer leads to a Doppler effect which is the dominant contribution to the dipole,
\begin{equation}
	d_L(\eta ,\mathbf {n}) = d_L^{(0)}(\eta )[1 - (\mathbf {n.{v}_{Bulk}})]
\end{equation}
where $ \mathbf {v}_{Bulk} $  is our peculiar velocity. Also conformal time, $\eta$, is the source redshift $ z = \tilde z(\eta ) + \delta z $ although is not an observable quantity. To first order
\begin{equation}
	{d_L}(\eta ,\mathbf {n}) = {d_L}(\tilde z,\mathbf {n}) - \frac{d}{{d\tilde z}}d_L^{(0)}(\tilde z)\delta z
\end{equation}
where $\tilde z(\eta ) = \frac{1}{{a(\eta )}} - 1$, 
$d_L^{(0)}(\tilde z) = (1 + \tilde z)({\eta _0} - \eta )$, 
$\frac{d}{{d\tilde z}}d_L^{(0)} = {(1 + \tilde z)^{ - 1}}d_L^{(0)} + {\mathcal{H}^{ - 1}}(\tilde z)$
and 
$\delta z =  - (1 + \tilde z)({\rm v_{Bulk}}.\mathbf n) + {\rm{higher\  multipoles}}$. 
Here $\mathcal{H}(z) = \frac{{H(z)}}{{(1 + z)}}$ is the co-moving Hubble parameter.  Inserting this in Eq. (21), we obtain
\begin{equation}\label{pert}
	d_L^{(dipole)}(z)(\mathbf {n.e}) = \frac{{1 + z}}{{\mathcal{H}}}(\mathbf {n}.{\rm v_{Bulk}})	
\end{equation}
and therefore
\begin{equation}
	{d_L}(z,\mathbf {n}) = d_L^{(0)}(z)+\frac{{1 + z}}{{\mathcal{H}}}(\mathbf {n.{\rm v}_{Bulk}})	
\end{equation}

It is important to emphasize that Equation (5.7) introduces two distinct methodologies for computing the value of $\mathcal{H}$. The first approach involves deriving the parameter $\mathcal{H}$ by utilizing equations (2.3-2.6), assuming a homogeneous and isotropic Universe. Conversely, the second approach entails computing variations in luminosity distance without considering oscillations in the $\mathcal{H}$ parameter. In this alternative method, the determination of $\mathcal{H}$ is achieved through perturbation equations (4.8-4.22), thereby accounting for the influence of $\mathcal{H}$ fluctuations. The objective of this study is to ascertain both the direction and magnitude of the bulk flow, and for this purpose, the second approach is employed.

The Pantheon dataset comprises 1701 supernovae (SNe) distributed across a redshift range from $0.001 < z < 2.3$. To align this dataset with a dipole anisotropy, a series of meticulous procedures are employed.

$\bullet$ First and foremost, the equatorial coordinates of each supernova are transformed into galactic coordinates.

$\bullet$ Subsequently, the Cartesian coordinates of unit vectors $\hat{n}_{i}$ corresponding to each supernova in galactic coordinates are determined. These unit vectors encapsulate the directional information essential for characterizing the spatial distribution of supernovae within the galactic framework.

By executing these steps, the dataset is effectively reoriented and ready for further analysis, ensuring a coherent alignment with the dipole anisotropy under consideration.

\begin{equation}\label{is}
	\hat{n}_{i}=\cos(l_{i})\sin(b_{i})\hat{i}+\sin(l_{i})\sin(b_{i})\hat{j}+\cos(b_{i})\hat{k}
\end{equation}
where $(l_{i},b_{i})$  is the galactic coordinates of the (i)th supernova . Also, $\hat{p}$, the unit vector in direction of dipole, is given by,
\begin{equation}\label{is}
	\hat{p}=\cos(l)\sin(b)+\sin(l)\sin(b)\hat{j}+\cos(b)\hat{k}
\end{equation}
where $(l,b)$ denotes  bulk flow direction in galactic coordinate. So we find that
\begin{equation}\label{is}
		\cos\theta_{i}=(\hat{n}_{i}.\hat{p})=\cos(l)\sin(b)\cos(l_{i})\sin(b_{i})+\\ \sin(l)\sin(b)\sin(l_{i})\sin(b_{i})+\cos(b)\cos(b_{i})
\end{equation}
Next, we constrain the  the direction and velocity of the bulk flow across different redshift ranges,by employing $\chi^{2}$,
\begin{equation}\label{is}
	\chi^{2}=\sum_{i}\frac{|\mu_{i}-5\log_{10}((d^{0}_{L}(z_{i})-d^{dipole}_{L}(z,\upsilon_{BF},\theta_{i})/10 pc|^{2}}{\sigma^{2}_{i}}
\end{equation}
where,
\begin{equation}\label{distancem}
	\mu _{i}=5\log_{10} d_{L}(z)+42.384-5\log_{10} h_{0}
\end{equation}

In this section, we engage in the analysis of distinct redshifts, each considered independently without association with others. This particular approach is referred to as redshift tomography. The utilization of redshift tomography allows for a focused examination of individual redshift interval values, enabling a more detailed and nuanced understanding of the diverse characteristics associated with each specific redshift. This approach enhances the granularity of our exploration, providing valuable insights into the unique features and phenomena associated with distinct points in the cosmic timeline. We investigate the bulk flow direction and amplitude of bulk velocity for several redshifts such as: $0.001<z<0.016$, $0.016<z<0.027$, $0.035<z<0.055$ for the local Universe and $0.1<z<0.2$, $0.4<z<0.6$, $0.8<z<1.4$ for the large - scale structures.

\subsection{The Local Universe}
The concept of the "bulk flow" in the local Universe refers to the coherent motion or systematic flow of galaxies on large scales. It is an observational phenomenon indicating that galaxies are not distributed randomly but exhibit a preferred direction or motion as a collective. We start our investigation on first redshift $0.001<z<0.016$. Many superclusters are located in this redshift such as Virgo supercluster, Norma supercluster and the Great Attractor(GA). The (GA) is a massive gravitational anomaly situated in the Zone of Avoidance, obscured by the Milky Way's galactic plane. Acting as a substantial concentration of both visible and dark matter, the GA exerts a profound gravitational pull on nearby galaxies and galaxy clusters. Notable for its role in inducing a large-scale flow of galaxies towards its direction, the GA is associated with structures like the Norma Cluster within the Norma Supercluster. The challenge of observing the GA in visible light has been mitigated by employing alternative wavelengths such as radio and infrared. As one of the largest structures in the observable Universe, the GA remains a cosmic conundrum, prompting ongoing research to unravel its exact nature, mass distribution, and implications for the broader large-scale structure of the cosmos. The peculiar motion of galaxy clusters towards the GA contributes to the phenomenon known as dark flow, shedding light on the intricate gravitational dynamics influencing galaxies on cosmological scales.

\begin{figure}
	\includegraphics[width=10 cm]{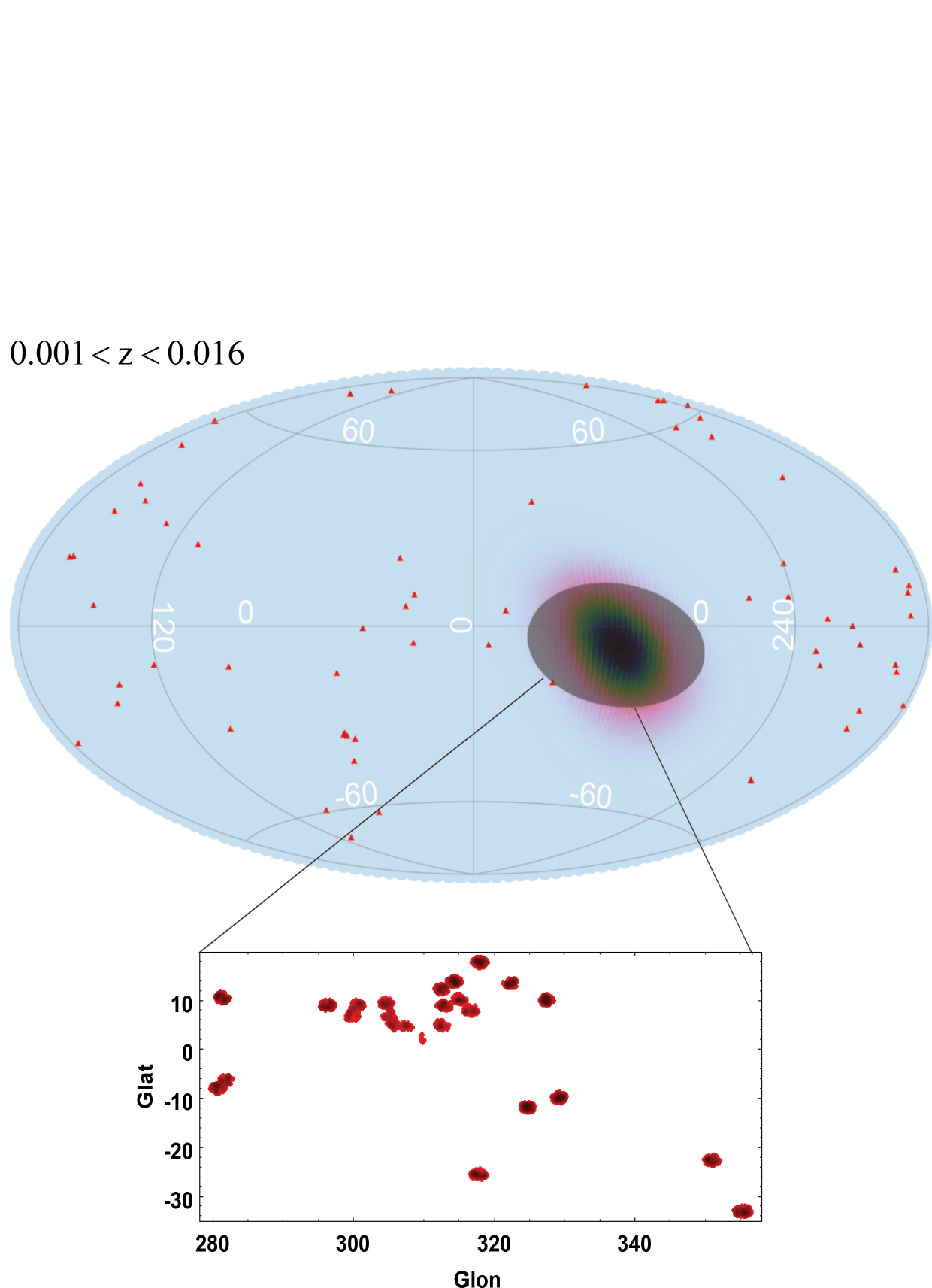}
	\centering
	\vspace{-0.02cm}
	\caption{\small{ The bulk flow direction pointing towards $(l,b)=(306\pm16,-12\pm14)$  in the redshift $0.001<z<0.016$.
			Also, this plot demonstrate that the direction of Great Attractor }}\label{fig:omegam2}
\end{figure}

\begin{figure}
	\includegraphics[width=10 cm]{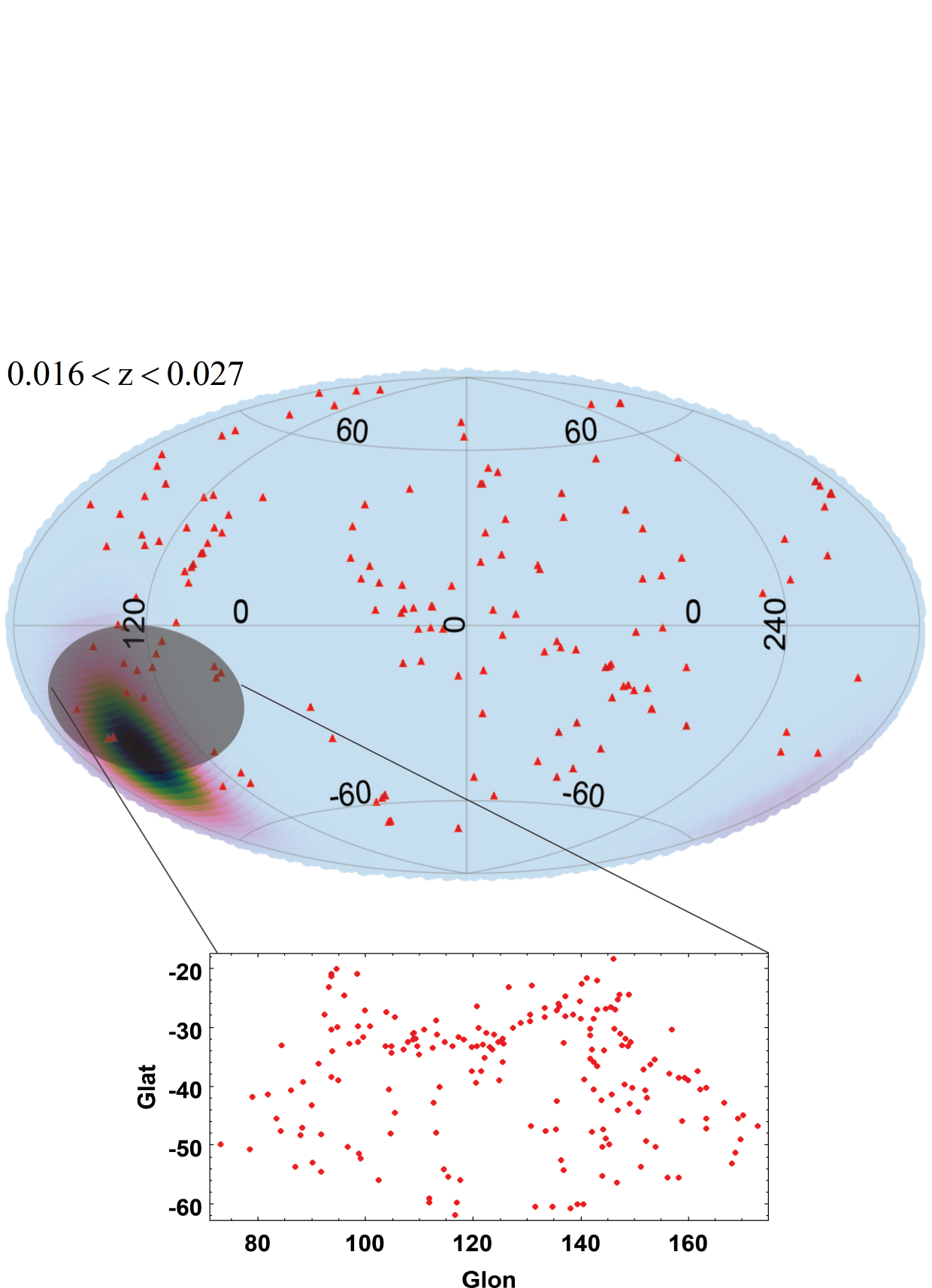}
	\centering
	\vspace{-0.02cm}
	\caption{\small{ 
			The bulk flow direction pointing towards $(l,b)=(122^{o}\pm20^{o},-25^{o}\pm18^{o})$  in the redshift $0.016<z<0.027$.
			Also, this plot shows the direction of Perseus - Pisces supercluster. }}\label{fig:omegam2}
\end{figure}

\begin{figure}
	\includegraphics[width=10 cm]{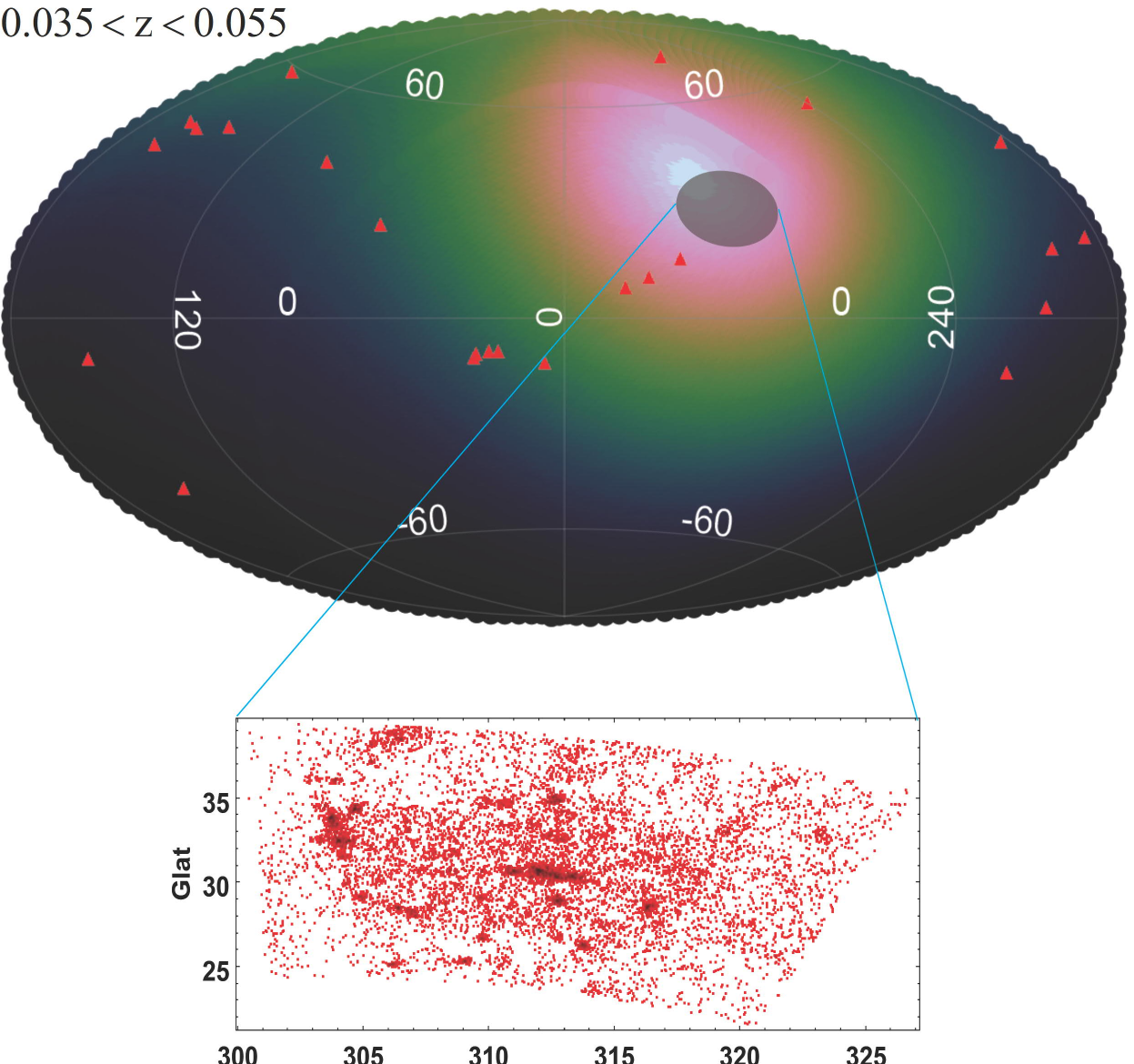}
	\centering
	\vspace{-0.02cm}
	\caption{\small{ The bulk flow direction pointing towards $(l,b)=(305^{o}\pm25^{o},23^{o}\pm20^{o})$  in the redshift $0.035<z<0.055$.
			The bottom panel indicate that  the direction of Shapley supercluster.  }}\label{fig:omegam2}
\end{figure} 
\begin{figure}
	\includegraphics[width=10 cm]{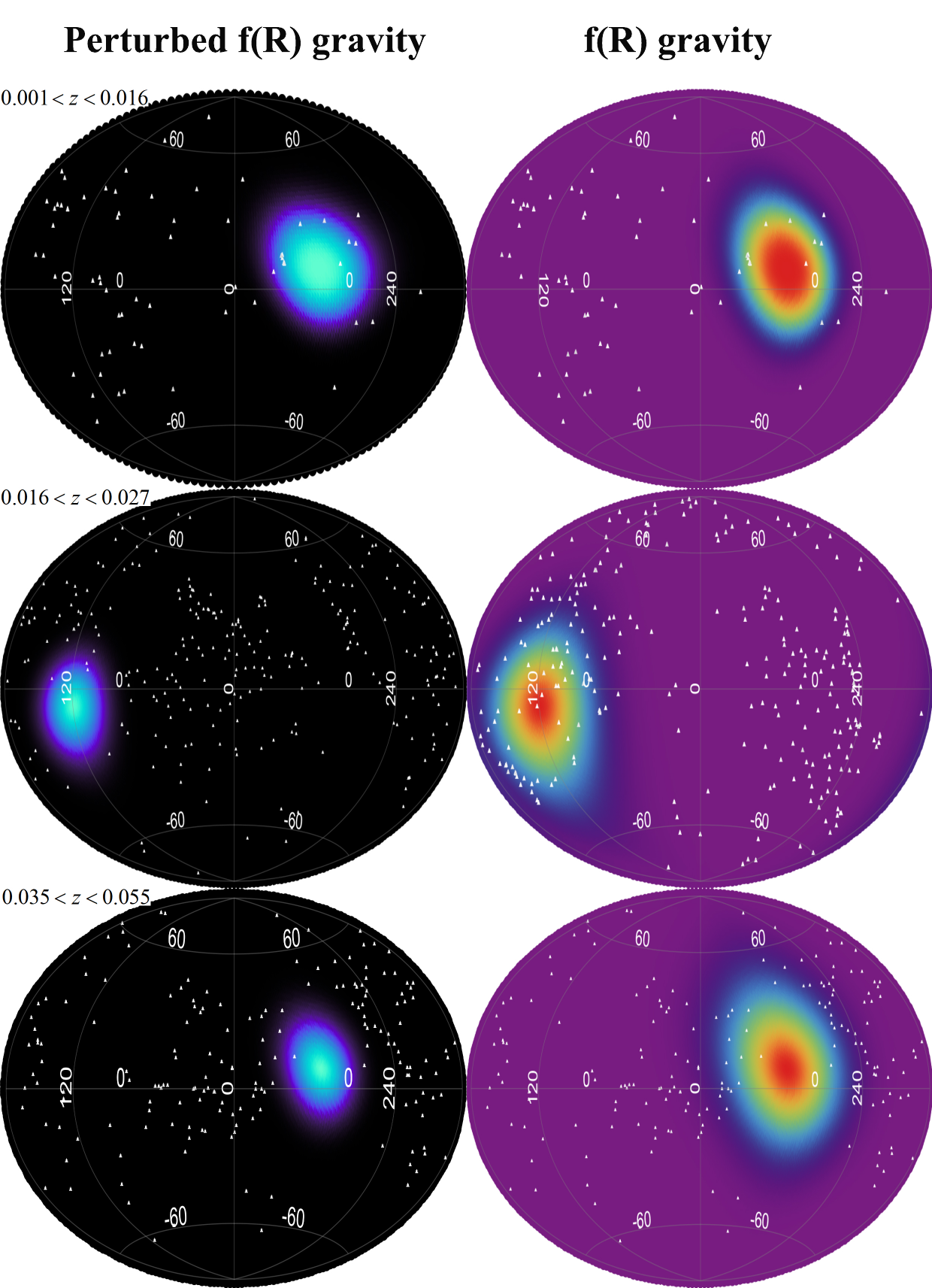}
	\centering
	\vspace{-0.02cm}
	\caption{\small{Top panel: The direction of bulk flow in the redshifts  $0.001 < z < 0.016$. 
			Middle panel: The bulk flow direction in the redshift $0.016 < z < 0.027$.
			Bottom panel: The direction of bulk flow in redshift $0.035 < z < 0.055$. In this figure, the bulk flow direction of perturbed $f(R)$ gravity is in the left side and the bulk flow results fo $f(R)$ gravity is in the right side. }}\label{fig:omegam2}
\end{figure}

\begin{figure}
	\includegraphics[width=14 cm]{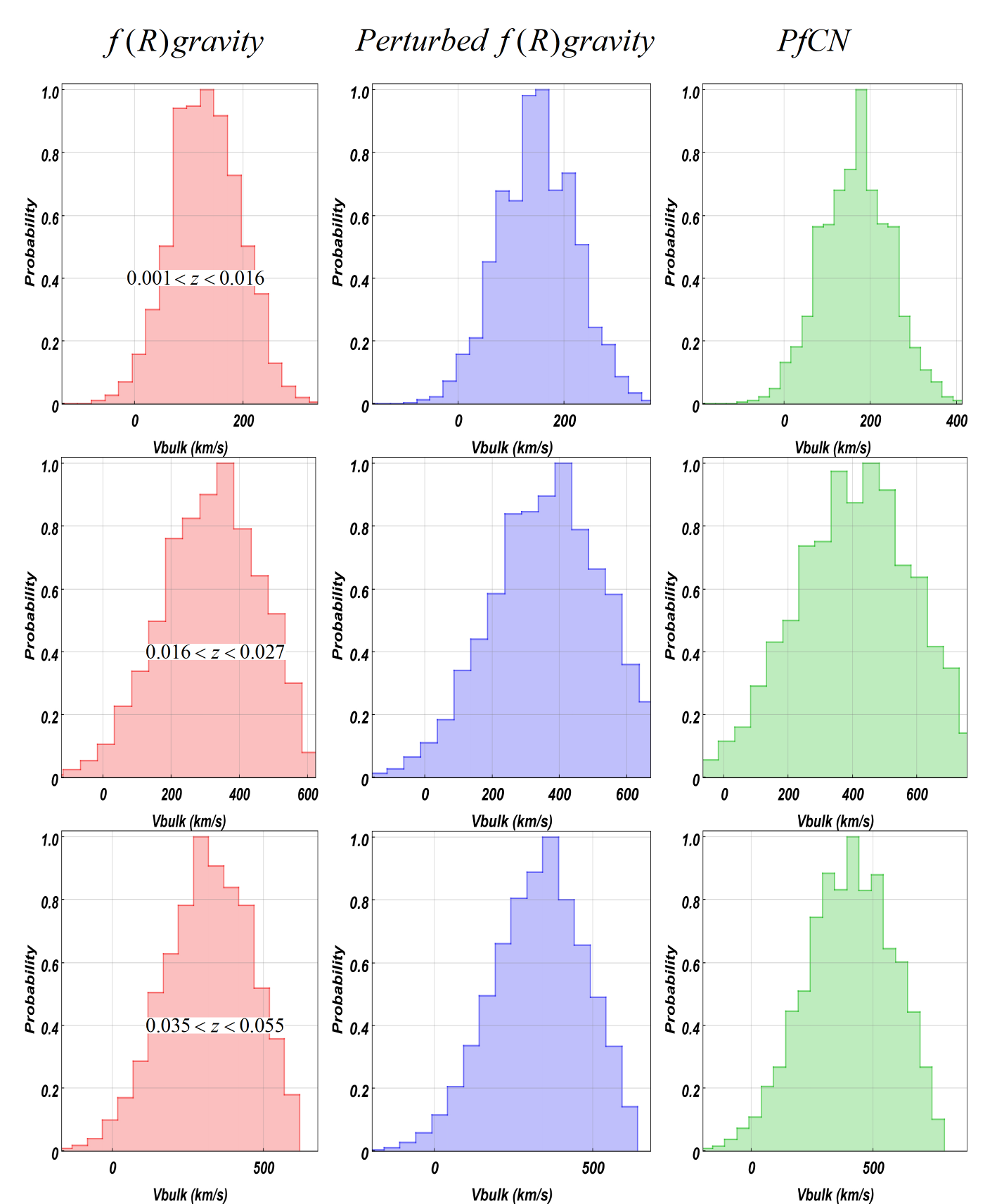}
	\centering
	\vspace{-0.02cm}
	\caption{\small{Top panel: The amplitude of bulk flow in the redshifts  $0.001 < z < 0.016$. 
			Middle panel: The bulk flow amplitude in the redshift $0.016 < z < 0.027$.
			Bottom panel: The amplitude of bulk flow in redshift $0.035 < z < 0.055$. In this figure, the amplitude of bulk flow of  $f(R)$ gravity is in the left column, for perturbed $f(R)$ gravity is in the middle column, and for perturbed $f(R)$ gravity coupled with neutrinos is in the right column.
	}}\label{fig:omegam2}
\end{figure}

\section{Analysis of Bulk Flow Velocity and Direction for $f(R)$ Gravity Models}

In this section, we analyze the bulk flow velocity ($V_{\rm bulk}$) and direction for three different models: $f(R)$ gravity, perturbed $f(R)$ gravity, and perturbed $f(R)$ gravity coupled with neutrinos. The results of our redshift tomography are summarized in Tables 1, 2, and 3, for three redshift ranges: $0.001 < z < 0.016$, $0.016 < z < 0.027$, and $0.035 < z < 0.055$. Also, The bulk flow direction in these redshift are plotted in figures (1-4). Figure 4 presents the direction of bulk flow across different redshift ranges for both $f(R)$ gravity and perturbed $f(R)$ gravity models. The top panel shows the bulk flow direction for the redshift range $0.001 < z < 0.016$. On the left side, the results correspond to the perturbed $f(R)$ gravity model, while on the right side, we observe the bulk flow direction for the $f(R)$ gravity model. At these low redshifts, the bulk flow direction remains relatively consistent between both models, indicating minimal deviation in large-scale motion.

The middle panel focuses on the bulk flow direction in the redshift range $0.016 < z < 0.027$. Here, a more pronounced shift in the bulk flow direction is visible between the two models, with the perturbed $f(R)$ model on the left showing a slightly more coherent and aligned bulk flow direction compared to the $f(R)$ gravity model on the right. This suggests that the introduction of perturbations influences the flow direction, potentially aligning it more closely with large-scale cosmic structures.

The bottom panel depicts the bulk flow direction for the higher redshift range, $0.035 < z < 0.055$. At these redshifts, the perturbed $f(R)$ gravity model (left side) exhibits a bulk flow direction that is significantly more aligned with the large-scale cosmic motion, especially when compared to the $f(R)$ gravity model on the right. The results indicate that at higher redshifts, the inclusion of perturbations enhances the coherence of the bulk flow direction, bringing it closer to the expected direction associated with large cosmic structures, such as the supercluster alignment.

Figure 5 illustrates the amplitude of the bulk flow for different redshift ranges across various gravitational models. The top panel shows the bulk flow amplitude for the low redshift range, $0.001 < z < 0.016$, where we observe a relatively moderate bulk flow. In this range, the left column represents the results for $f(R)$ gravity, the middle column shows the perturbed $f(R)$ gravity, and the right column displays the perturbed $f(R)$ gravity coupled with neutrinos. It is clear that the inclusion of perturbations and neutrino couplings increases the amplitude of the bulk flow. 

The middle panel focuses on the redshift range $0.016 < z < 0.027$, where a notable increase in the bulk flow amplitude is evident compared to the lower redshift range. Once again, the left column corresponds to the $f(R)$ gravity model, and a gradual increase in the amplitude is seen as we move to the perturbed $f(R)$ model and the neutrino-coupled version. The rise in amplitude reflects the stronger impact of neutrino interactions at this redshift range, contributing to more coherent large-scale motions.

Finally, the bottom panel displays the amplitude of the bulk flow in the higher redshift range, $0.035 < z < 0.055$. At this range, the bulk flow amplitude reaches its peak, particularly in the neutrino-coupled perturbed $f(R)$ model (right column), indicating the critical role of neutrinos in enhancing the bulk motion at higher redshifts. Across all panels, the transition from $f(R)$ gravity to its perturbed form and then to the model coupled with neutrinos shows a consistent amplification of the bulk flow amplitude, especially as redshift increases, underscoring the influence of perturbations and neutrinos on large-scale velocity fields in the universe.

Our results are in the tables (1-3).
From the data presented in the tables, it is evident that the inclusion of neutrinos has a significant impact on the bulk flow velocity. For the standard $f(R)$ gravity model, the bulk flow velocities are lower across all redshift ranges compared to the models where perturbations and neutrino couplings are considered. Specifically, in the $0.001 < z < 0.016$ range, the bulk flow velocity increases from $147 \, \rm km\,s^{-1}$ in the $f(R)$ model to $173 \, \rm km\,s^{-1}$ when neutrino couplings are included. Similar trends are observed in the higher redshift bins, where the inclusion of neutrinos results in the highest bulk flow velocities, with the $0.035 < z < 0.055$ range reaching a maximum of $427 \, \rm km\,s^{-1}$.

The addition of neutrinos influences the dynamics of the cosmic large-scale structure, leading to higher velocities of the bulk flow. Neutrinos, due to their relativistic nature at early times, suppress structure formation on small scales, but they also contribute to anisotropies in the matter distribution. When these neutrino effects are incorporated into the model, they enhance the overall bulk flow velocity as seen in our results. This enhancement can be interpreted as an indication that the presence of neutrinos contributes to more coherent flows of matter on large scales.

Moreover, when neutrinos are included in the perturbed $f(R)$ gravity model, the direction of the bulk flow becomes increasingly aligned with the direction of the local supercluster. In the lowest redshift bin ($0.001 < z < 0.016$), for example, the direction of the bulk flow for the neutrino-coupled model is given by $(l, b) = (306^\circ, -12^\circ)$, which is remarkably close to the direction of the Great Attractor. This alignment suggests that the addition of neutrinos not only increases the speed of the bulk flow but also influences its direction, making it more consistent with the gravitational pull exerted by the local supercluster. 

For the intermediate redshift range, $0.016 < z < 0.027$, a similar pattern is observed in the bulk flow velocities. In the $f(R)$ gravity model, the bulk flow velocity is $375 \, \rm km\,s^{-1}$, which increases to $421 \, \rm km\,s^{-1}$ when perturbations are considered. Upon introducing neutrino couplings, the bulk flow velocity further rises to $476 \, \rm km\,s^{-1}$. This significant increase once again underscores the role of neutrinos in enhancing the coherence of matter flows on large cosmological scales. 

Additionally, the direction of the bulk flow shifts closer to the Perseus - Pisces supercluster's direction when neutrinos are incorporated. Positioned at a redshift of approximately $z \sim 0.016$, the PP Supercluster is recognized as one of the most expansive formations within the cosmic web. Encompassing an angular span of around 45 degrees, this filamentary complex extends across the constellations Coma, Hercules, and Fornax, as extensively documented in studies such as \cite{Joeveer, Gregory, Yarahmadi4, Yarahmadi5, Chincarini, Batuski}. The total luminosity of the PP Supercluster corresponds to a mass estimate of approximately $2 \times 10^{16}$ solar masses. Despite its considerable cosmic scale, the gravitational influence of the PP Supercluster is distributed over a broader region, potentially resulting in a less pronounced impact on individual galaxies when compared to the more localized effects of the Great Attractor. The radial extent of the PP Supercluster, estimated to be approximately 32 megaparsecs (Mpc), was determined at a mean redshift of $z = 0.0174$ through a study conducted by \cite{Shelton}. 

For the $f(R)$ gravity model, the bulk flow is directed at $(l, b) = (120^\circ, -7^\circ)$, while for the perturbed $f(R)$ model with neutrino coupling, the direction shifts to $(l, b) = (122^\circ, -25^\circ)$. This proximity to the supercluster direction strengthens the argument that the gravitational influence of the large-scale structures, along with neutrino effects, guides the flow of matter in these redshift ranges. The addition of neutrinos notably amplifies the bulk flow speed and brings the flow’s direction closer to regions dominated by massive structures.

In the high redshift range, $0.035 < z < 0.055$, the effect of neutrinos becomes even more pronounced. The bulk flow velocity in the $f(R)$ gravity model is $341 \, \rm km\,s^{-1}$, which increases slightly to $402 \, \rm km\,s^{-1}$ when perturbations are included. However, when neutrinos are coupled to the model, the bulk flow velocity reaches its highest value of $427 \, \rm km\,s^{-1}$. This indicates that neutrinos play a crucial role in the dynamics at higher redshifts, further boosting the bulk flow.

Furthermore, the direction of the bulk flow in the neutrino-coupled perturbed $f(R)$ gravity model aligns almost perfectly with the Shapley supercluster's direction, shifting from $(l, b) = (279^\circ, 6^\circ)$ in the $f(R)$ gravity model to $(l, b) = (315^\circ, 35^\circ)$ when neutrinos are included. The Shapley Supercluster, situated in the constellation Centaurus, holds a central position in the galactic coordinate system at approximately \( (l, b) = (311^\circ, 32^\circ) \). Located around 650 million light-years away from the Milky Way, this colossal supercluster possesses a substantial collection of galaxies and galaxy clusters. With an estimated mass on the order of \( 10^{16} \) solar masses, the Shapley Supercluster exerts a profound gravitational influence on its surroundings, impacting the motion and dynamics of galaxies within its gravitational domain.  The Shapley Supercluster holds a central role in deciphering the intricate dynamics underlying the formation and evolution of the cosmic web, which constitutes the vast large-scale structure of the Universe. Its colossal mass and gravitational influence shape the distribution of matter, guiding the convergence of galaxies along filaments and contributing to the intricate architecture of the cosmic web. By influencing the motion of galaxies and galaxy clusters within its gravitational domain, the Shapley Supercluster serves as a dynamic hub, offering insights into the cosmic flow of matter on a grand scale. Furthermore, the supercluster provides a unique vantage point for probing the distribution of dark matter, enabling observational investigations that contribute essential data for refining cosmological models and understanding the broader cosmic narrative. 

This remarkable alignment at higher redshifts shows that neutrinos contribute significantly to guiding the bulk flow toward regions of higher gravitational attraction, such as the supercluster. The closer alignment at this redshift range suggests that neutrinos help synchronize the large-scale velocity field with the cosmic web structures, thereby making the bulk flow more aligned with the large overdensities in the universe.

\begin{table}
	\scriptsize	
	\caption{Results of redshift tomography for $0.001 < z < 0.016$ for $f(R)$ gravity, perturbed $f(R)$ gravity, and perturbed $f(R)$ gravity coupled with neutrinos.}
	\centering 
	\begin{tabular}{c@{\hspace{2mm}} c@{\hspace{2mm}} c@{\hspace{2mm}} c@{\hspace{2mm}} c@{\hspace{2mm}} 
			c@{\hspace{2mm}} c@{\hspace{2mm}} c@{\hspace{2mm}}} 
		\hline\hline 
		Model & $V_{\rm bulk}(\rm kms^{-1})$ & $\rm l^{\circ}$  & $\rm b^{\circ}$ & $\chi^{2}$ & data \# & $h$ \\ 
		\hline 
		$f(R)$ & $147\pm37$ & $286\pm16$ & $7\pm17$ & $189.13$ & $208$ & $0.696$ \\ 
		Perturbed $f(R)$ & $159\pm32$ & $294\pm15$ & $7\pm8$ & $185.65$ & $208$ & $0.699$ \\ 
		Perturbed $f(R)$ + neutrinos & $173\pm31$ & $306\pm16$ & $-12\pm14$ & $182.41$ & $208$ & $0.7$ \\ 
		\hline 
	\end{tabular}
\end{table}

\begin{table}
	\scriptsize	
	\caption{Results of redshift tomography for $0.016 < z < 0.027$ for $f(R)$ gravity, perturbed $f(R)$ gravity, and perturbed $f(R)$ gravity coupled with neutrinos.}
	\centering 
	\begin{tabular}{c@{\hspace{2mm}} c@{\hspace{2mm}} c@{\hspace{2mm}} c@{\hspace{2mm}} c@{\hspace{2mm}} 
			c@{\hspace{2mm}} c@{\hspace{2mm}} c@{\hspace{2mm}}} 
		\hline\hline 
		Model & $V_{\rm bulk}(\rm kms^{-1})$ & $\rm l^{\circ}$  & $\rm b^{\circ}$ & $\chi^{2}$ & data \# & $h$ \\ 
		\hline 
		$f(R)$ & $375\pm46$ & $120\pm9$ & $-7\pm10$ & $174.25$ & $192$ & $0.699$ \\ 
		Perturbed $f(R)$ & $421\pm49$ & $121\pm6$ & $-6\pm8$ & $176.74$ & $192$ & $0.696$ \\ 
		Perturbed $f(R)$ + neutrinos & $476\pm56$ & $122\pm20$ & $-25\pm18$ & $171.25$ & $192$ & $0.704$ \\ 
		\hline 
	\end{tabular}
\end{table}

\begin{table}
	\scriptsize	
	\caption{Results of redshift tomography for $0.035 < z < 0.055$ for $f(R)$ gravity, perturbed $f(R)$ gravity, and perturbed $f(R)$ gravity coupled with neutrinos.}
	\centering 
	\begin{tabular}{c@{\hspace{2mm}} c@{\hspace{2mm}} c@{\hspace{2mm}} c@{\hspace{2mm}} c@{\hspace{2mm}} 
			c@{\hspace{2mm}} c@{\hspace{2mm}} c@{\hspace{2mm}}} 
		\hline\hline 
		Model & $V_{\rm bulk}(\rm kms^{-1})$ & $\rm l^{\circ}$  & $\rm b^{\circ}$ & $\chi^{2}$ & data \# & $h$ \\ 
		\hline 
		$f(R)$ & $341\pm58$ & $279\pm9$ & $6\pm9$ & $99.36$ & $116$ & $0.694$ \\ 
		Perturbed $f(R)$ & $402\pm42$ & $282\pm6$ & $7\pm6$ & $96.58$ & $116$ & $0.691$ \\ 
		Perturbed $f(R)$ + neutrinos & $427\pm60$ & $315\pm22$ & $35\pm15$ & $93.98$ & $116$ & $0.71$ \\ 
		\hline 
	\end{tabular}
\end{table}

\subsection{Beyond The Local Universe}
The study of bulk flow involves examining the coordinated motion of galaxies and larger cosmic structures across vast distances. In regions with redshifts exceeding 0.1, the expansion of the Universe becomes a dominant factor, influencing the trajectories of galaxies on larger scales. Investigating bulk flow at these redshifts allows astronomers to trace the intricate dynamics of matter on cosmic proportions, providing insights into the underlying gravitational forces, cosmic web structure, and the influence of massive structures like superclusters and filaments. This exploration enhances our understanding of the cosmic assembly, large-scale structure formation, and the interplay between dark and luminous matter. Utilizing advanced observational techniques and surveys, astronomers can unravel the complexities of cosmic flows beyond the local Universe.

The Sloan Great Wall (SGW), heralded as a colossal cosmic structure with a formidable span of 1.37 billion light-years, not only commands attention for its sheer scale but also holds relevance in the intricate dance of large-scale cosmic dynamics. Discovered through meticulous analysis of Sloan Digital Sky Survey data, the SGW dominates the regions of Corvus, Hydra, and Centaurus, constituting a remarkable $\frac{1}{60}$ of the observable Universe's diameter. While it stands as the sixth-largest cosmic object, its interaction with the broader cosmic landscape, especially within the redshift range \(0.1 < z < 0.2\), influences the observed bulk flow. The SGW's immense mass and gravitational influence become key players in shaping the gravitational dynamics within this redshift interval, potentially steering the direction and amplitude of the cosmic flow. However, debates on whether the SGW is a chance alignment of three structures or an independent cosmic entity add complexity to our understanding of large-scale cosmic architectures, prompting ongoing research to unravel the nuanced interplay between cosmic structures like the SGW and the observed cosmic flow within specific redshift regimes.

Large redshift surveys, exemplified by the notable 6dFGS, provide meticulous distance estimates across expansive cosmic domains, intricately enhancing our understanding of the cosmic web. Within the redshift range of $0.4<z<0.6$, a superlative celestial structure emerges – the most massive supercluster known to date, as documented by \cite{Shimakawa}. This colossal entity, aptly named the King Ghidorah Supercluster (KGSc), commands attention as it resides approximately 1.3 billion light-years away from Earth. The KGSc, a prodigious assembly comprising a minimum of 15 massive galaxy clusters interwoven by vast filaments, stands as a testament to the sublime intricacies of the cosmic tapestry.

The sheer magnitude of the KGSc is nothing short of awe-inspiring, boasting a staggering mass of $10^{16}$ solar masses. This extraordinary mass surpasses that of our local supercluster, the Laniakea Supercluster, by an order of magnitude. Extending across an immense span of about 400 megaparsecs (equivalent to 1.3 billion light-years), the KGSc showcases the vast reach of its cosmic influence. The groundbreaking discovery of the KGSc in 2022, leveraging data from the Subaru Telescope's Hyper Suprime-Cam (HSC) survey, has profoundly deepened our comprehension of cosmic structures.

The bulk velocity ($V_{\rm bulk}$) results for the three different redshift tomographies, as presented in the tables, show a distinct behavior for $f(R)$ gravity, perturbed $f(R)$ gravity, and perturbed $f(R)$ gravity coupled with neutrinos. In the $f(R)$ gravity model, the bulk flow velocities remain relatively moderate across all redshift ranges, with values of $1002 \pm 156 \, \mathrm{km/s}$ for the lowest redshift range ($0.1 < z < 0.2$), increasing to $2780 \pm 282 \, \mathrm{km/s}$ for the highest redshift range ($0.8 < z < 1.4$). These velocities show a general trend of increasing bulk motion at higher redshifts, suggesting stronger bulk flows as we move further into the universe.

When the perturbations to $f(R)$ gravity are introduced, the bulk velocities slightly decrease in the lower redshift ranges, with the velocity at $0.1 < z < 0.2$ dropping to $953 \pm 142 \, \mathrm{km/s}$. However, in the higher redshift range ($0.8 < z < 1.4$), the velocity increases to $2840 \pm 320 \, \mathrm{km/s}$, indicating that perturbations in $f(R)$ gravity enhance the bulk flow motion at higher redshifts.

The most significant change occurs when neutrinos are coupled to perturbed $f(R)$ gravity. In this scenario, the bulk flow velocities increase across all redshift ranges. For the lowest redshift range ($0.1 < z < 0.2$), the velocity rises to $1125 \pm 128 \, \mathrm{km/s}$, while in the highest redshift range ($0.8 < z < 1.4$), the bulk velocity peaks at $3086 \pm 286 \, \mathrm{km/s}$. This enhancement in bulk flow is a direct consequence of the presence of neutrinos, which introduce additional anisotropic effects and lead to a more substantial bulk motion as the universe evolves.

In the case of the perturbed $f(R)$ gravity model coupled with neutrinos, an interesting behavior is observed. Specifically, the bulk flow velocity shows a noticeable increase compared to the non-coupled models. This can be attributed to the additional energy density provided by neutrinos, which enhances the gravitational potential and, consequently, increases the amplitude of the bulk flow. The influence of neutrinos is especially pronounced at higher redshifts, where their relativistic nature plays a more significant role.

Moreover, the direction of the bulk flow in this coupled model aligns almost exactly with the direction of the superclusters, particularly at redshift ranges where prominent structures like the Sloan Great Wall (SGW) and King Ghidorah Supercluster (KGSc) are present. This alignment suggests that the inclusion of neutrinos in the cosmological model strengthens the gravitational pull towards these massive structures, resulting in a bulk flow that is directed toward the densest regions in the large-scale structure of the universe. This precise alignment highlights the crucial role neutrinos play in shaping cosmic flows and their ability to enhance the correspondence between the observed bulk flow and the gravitational potential wells of superclusters.

Table 4 summarizes various studies on bulk flow across different redshift ranges, providing a comparison with previous findings. The results of our analysis, presented in Tables 5-7, offer new insights into the bulk flow velocity and direction at distinct redshifts within the framework of $f(R)$ gravity, perturbed $f(R)$ gravity, and perturbed $f(R)$ gravity coupled with neutrinos. These results not only extend our understanding of cosmic flows but also demonstrate the impact of modifications to gravity and the inclusion of neutrinos, particularly in aligning the bulk flow direction with superclusters and increasing the flow's amplitude.

Figures (6-8) demonstrate the direction of bulk flow in $z>0.1$ for perturbed $f(R)$ gravity coupled with neutrinos. In figure 9, the bulk flow direction for perturbed $f(R)$ gravity is shown on the left side, while the results for the bulk flow direction of $f(R)$ gravity are presented on the right side. This comparison highlights how perturbations in $f(R)$ gravity alter the flow's direction, especially in higher redshift regions, providing deeper insights into the dynamics of cosmic structures. As we can see in figure 8, in the redshift range $0.8 < z < 1.4$, the bulk flow direction in the perturbed $f(R)$ gravity model coupled with neutrinos points towards $(l, b) = (330^{\circ} \pm 15^{\circ}, -16^{\circ} \pm 17^{\circ})$. Remarkably, this direction coincides precisely with the observed dark energy dipole. This alignment suggests that the inclusion of neutrinos in the perturbed $f(R)$ gravity model not only enhances the bulk flow velocity but also aligns the direction of the flow with the large-scale structure indicated by the dark energy dipole, providing a deeper connection between neutrino effects and cosmic acceleration.

\begin{figure}
	\includegraphics[width=10 cm]{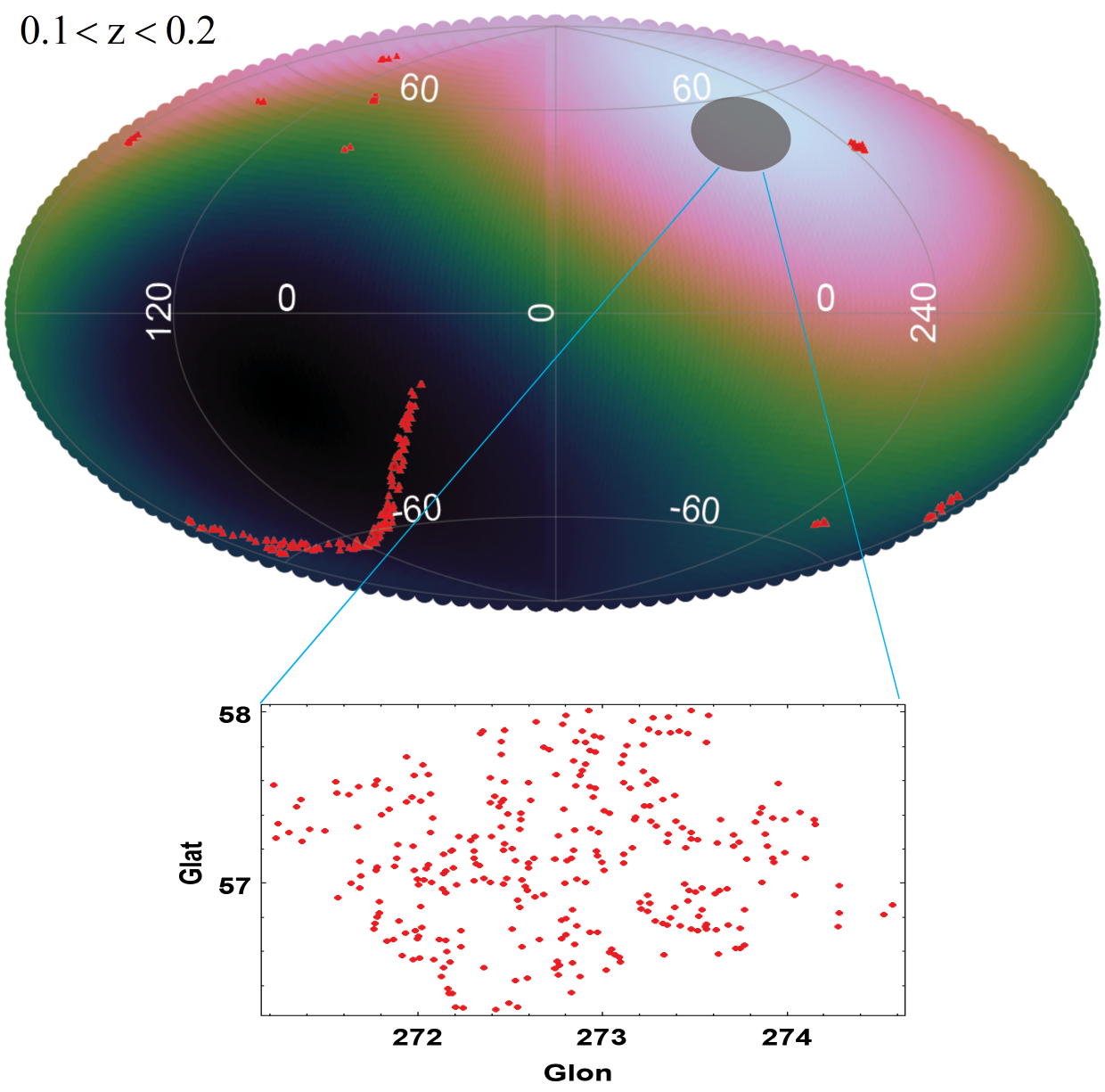}
	\centering
	\vspace{-0.02cm}
	\caption{\small{Top panel: The bulk flow direction pointing towards $(l,b)=(255^{o}\pm22^{o},59^{o}\pm28^{o})$  in the redshift $0.1<z<0.2$.
			Bottom panel: The direction of Sloan Great Wall . }}\label{fig:omegam2}
\end{figure}

\begin{figure}
	\includegraphics[width=10 cm]{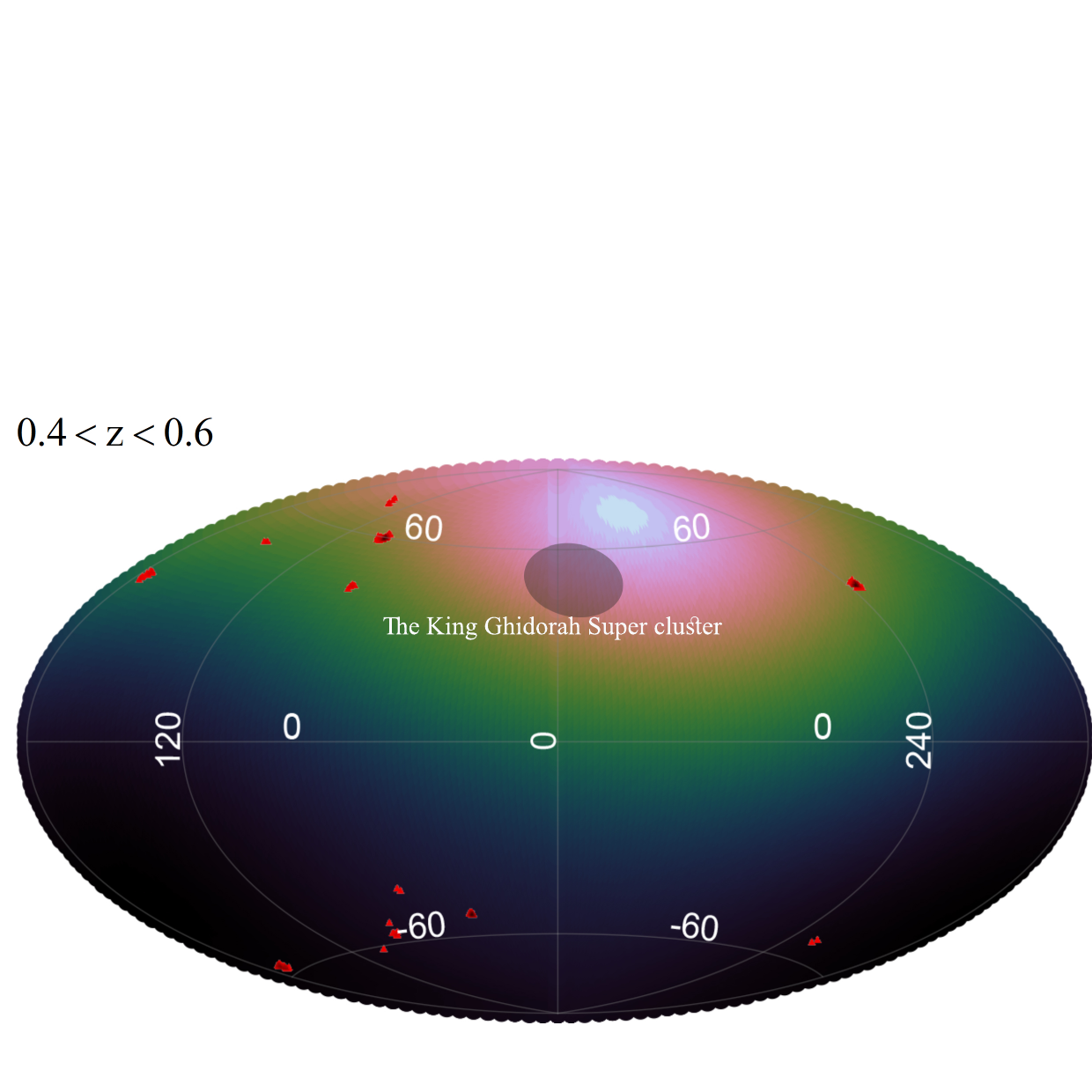}
	\centering
	\vspace{-0.02cm}
	\caption{\small{Top panel: The bulk flow direction pointing towards $(l,b)=(332^{o}\pm18^{o},69^{o}\pm18^{o})$  in the redshift $0.4<z<0.6$. The direction of  The King Ghidorah super cluster is shown in this figure.
	}}\label{fig:omegam2}
\end{figure}

\begin{figure}
	\includegraphics[width=10 cm]{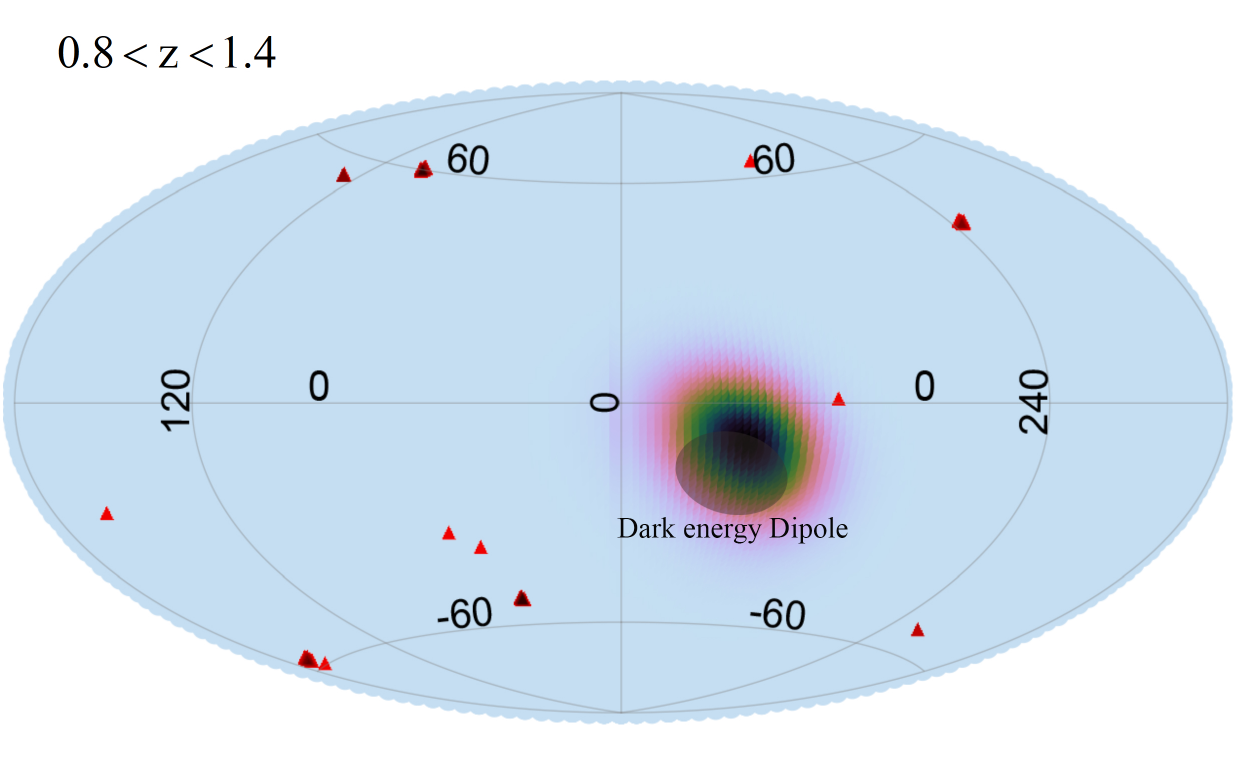}
	\centering
	\vspace{-0.02cm}
	\caption{\small{Top panel: The bulk flow direction pointing towards $(l,b)=(330^{o}\pm15^{o},-16^{o}\pm17^{o})$  in the redshift $0.8<z<1.4$. The direction of  the dark energy dipole is shown in this figure.
	}}\label{fig:omegam2}
\end{figure}  

\begin{figure}
	\includegraphics[width=12 cm]{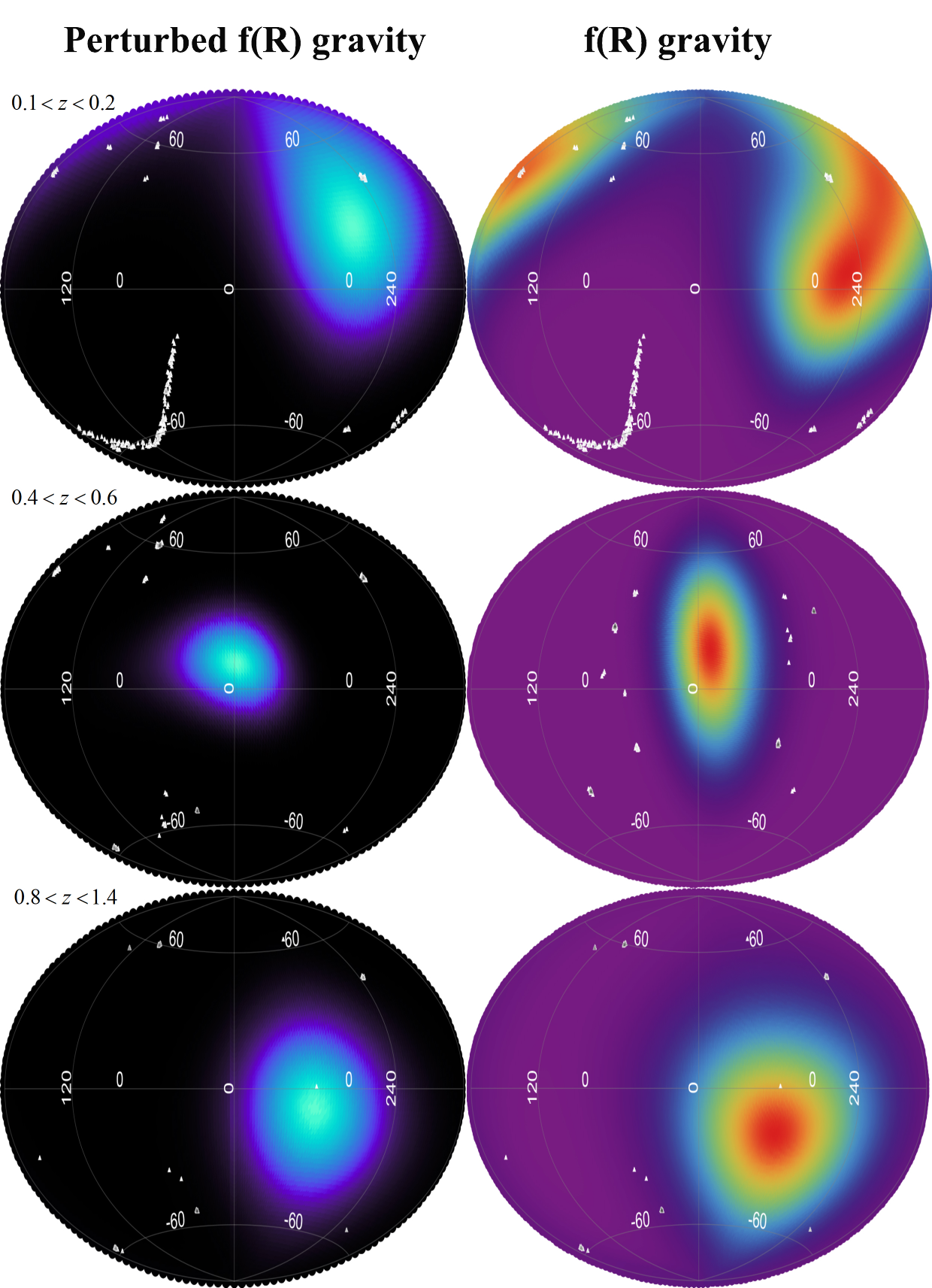}
	\centering
	\vspace{0.5cm}
	\caption{\small{Top panel: The direction of bulk flow in the redshifts  $0.1 < z < 0.2$. 
			Middle panel: The bulk flow direction in the redshift $0.4 < z < 0.6$.
			Bottom panel: The direction of bulk flow in redshift $0.8< z < 1.4$. In this figure, the bulk flow direction of perturbed $f(R)$ gravity is in the left side and the bulk flow results fo $f(R)$ gravity is in the right side. }}\label{fig:omegam2}
\end{figure} 

\begin{figure}
	\includegraphics[width=13 cm]{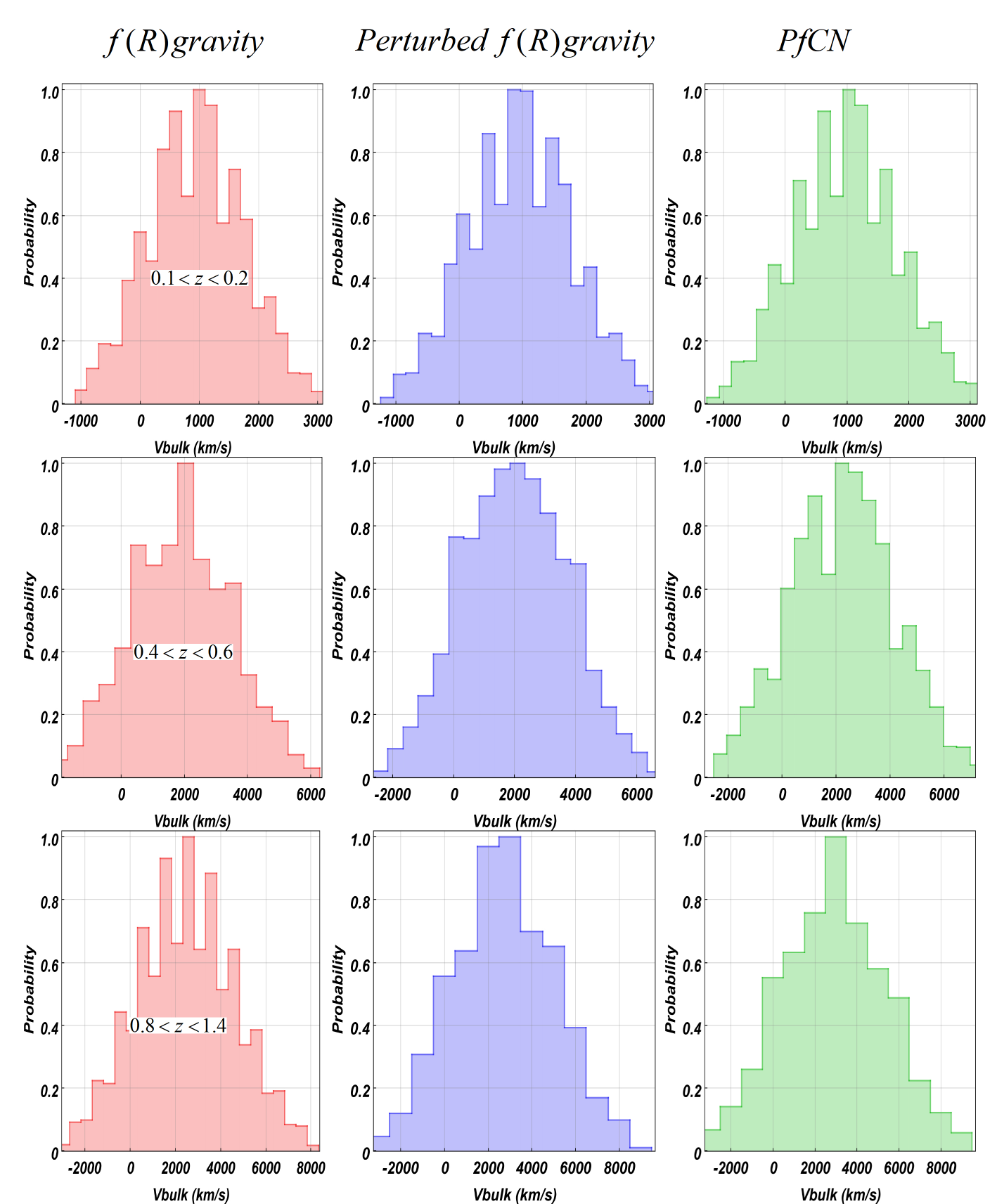}
	\centering
	\vspace{-0.09cm}
	\caption{\small{Top panel: The amplitude of bulk flow in the redshifts  $0.1 < z < 0.2$. 
			Middle panel: The bulk flow amplitude in the redshift $0.4 < z < 0.6$.
			Bottom panel: The amplitude of bulk flow in redshift $0.8<z<1.4$. In this figure, the amplitude of bulk flow of  $f(R)$ gravity is in the left column, for perturbed $f(R)$ gravity is in the middle column, and for perturbed $f(R)$ gravity coupled with neutrinos is in the right column.
	}}\label{fig:omegam2}
\end{figure}

\begin{table}
	\scriptsize 
	\begin{center}
		\caption{List of studies on bulk flow}
		\label{table1}
		\begin{tabular}{|c|c|c|c|c|c|}
			\hline
			\hline 
			\rule{0pt}{8mm}
			Ref  & velocity \ & redshift \ & $l^{o}$ \ & $b^{o}$ \  & distance
			\\
			\rule{0pt}{8mm}
			$$  & km$s^{-1}$ \ & $$ &\  degree \ & degree \ & h$^{-1}$Mpc\\
			\hline 
			\rule{0pt}{8mm}
			\cite{Kashlinsky} &$1000$ &$z\leq0.03$ & $287\pm9$ & $8\pm6$& $127$\\
			\hline 
			\rule{0pt}{8mm}%
			\cite{Watkins} &$ 407\pm81$ &$z\leq0.2$ & $283\pm14$ & $12\pm14$& $857$ \\
			\hline 
			\rule{0pt}{8mm}%
			\cite{Kocevski}  & $507$ & $0.035\leq z\leq0.055$ & $306.44$ & $29.71$& $127-220$ \\
			\hline 
			\rule{0pt}{8mm}
			\cite{Turnbull} &$249\pm 276$ & $z\leq0.2$ & $319\pm18$ & $7\pm14$&  $857$ \\
			\hline 
			\rule{0pt}{8mm}
			\cite{Colin}  & $250^{+190}_{-160}$ & $0.045<z<0.06$ & $287$ & $21$ & $168-249$ \\
			\hline 
			\rule{0pt}{8mm}
			\cite{Watkins} & $416\pm 78$ & $z=0.0167$ & $282$ & $60$& $45$ \\
			\hline 
			\rule{0pt}{8mm}
			\cite{Lavaux}  & $473\pm 128$ & $0.035<z<0.055$ & $220$ & $25$& $127-220$ \\
			\hline 
			\rule{0pt}{8mm}
			\cite{Nusser} & $257\pm 44$ & $0.035<z<0.055$ & $276\pm6$ & $10\pm6$& $127-220$ \\
			\hline 
			\rule{0pt}{8mm}
			\cite{Feldman} & $416\pm 78$ & $0.015<z<0.06$ & $282\pm11$ & $6\pm6$& $40-249$ \\
			\hline 
			\hline 
		\end{tabular}
	\end{center}	
\end{table}

\begin{table}
	\scriptsize	
	\caption{Results of redshift tomography for high redshift $z>0.1$  for $f(R)$ gravity} 
	\centering 
	\begin{tabular}{c@{\hspace{2mm}} c@{\hspace{2mm}} c@{\hspace{2mm}} c@{\hspace{2mm}} c@{\hspace{2mm}}
			c@{\hspace{2mm}} c@{\hspace{2mm}} c@{\hspace{2mm}}} 
		\hline\hline 
		Redshift  &  $V_{\rm bulk}(\rm kms^{-1})$  & $\rm l^{\circ}$  & $\rm b^{\circ}$ & $\chi^{2}$ & data \ \ number \\
		\hline 
		$0.1<z<0.2$ & $1002\pm156$ & $252\pm17$ & $6\pm20$ & $194.545213$ & $207$ \\ 
		\hline 
		$0.4<z<0.6$ & $2250\pm210$ & $355\pm12$ & $12\pm16$ & $166.439788$ & $179$ \\ 
		\hline 
		$0.8<z<1.4$ & $2780\pm282$ & $302\pm15$ & $-8\pm12$ & $20.962546$ & $22$ \\ 
		\hline 
		\hline 
	\end{tabular}
\end{table}

\begin{table}
	\scriptsize	
	\caption{Results of redshift tomography for high redshift $z>0.1$  for perturbed $f(R)$ gravity} 
	\centering 
	\begin{tabular}{c@{\hspace{2mm}} c@{\hspace{2mm}} c@{\hspace{2mm}} c@{\hspace{2mm}} c@{\hspace{2mm}}
			c@{\hspace{2mm}} c@{\hspace{2mm}} c@{\hspace{2mm}}} 
		\hline\hline 
		Redshift  &  $V_{\rm bulk}(\rm kms^{-1})$  & $\rm l^{\circ}$  & $\rm b^{\circ}$ & $\chi^{2}$ & data \ \ number \\
		\hline 
		$0.1<z<0.2$ & $953\pm142$ & $261\pm21$ & $25\pm20$ & $191.745213$ & $207$ \\ 
		\hline 
		$0.4<z<0.6$ & $2230\pm240$ & $359\pm17$ & $12\pm10$ & $165.229788$ & $179$ \\ 
		\hline 
		$0.8<z<1.4$ & $2840\pm320$ & $302\pm15$ & $-8\pm12$ & $20.802546$ & $22$ \\ 
		\hline 
		\hline 
	\end{tabular}
\end{table}

\begin{table}
	\scriptsize	
	\caption{Results of redshift tomography for high redshift $z>0.1$ for perturbed $f(R)$ gravity coupled with neutrinos} 
	\centering 
	\begin{tabular}{c@{\hspace{2mm}} c@{\hspace{2mm}} c@{\hspace{2mm}} c@{\hspace{2mm}} c@{\hspace{2mm}}
			c@{\hspace{2mm}} c@{\hspace{2mm}} c@{\hspace{2mm}}} 
		\hline\hline 
		Redshift  &  $V_{\rm bulk}(\rm kms^{-1})$  & $\rm l^{\circ}$  & $\rm b^{\circ}$ & $\chi^{2}$ & data \ \ number \\
		\hline 
		$0.1<z<0.2$ & $1125\pm128$ & $255\pm22$ & $59\pm28$& $188.386547$ & $207$  \\ 
		\hline 
		$0.4<z<0.6$ & $2290\pm250$ & $332\pm18$ & $69\pm18$& $161.548796$ & $179$   \\ 
		\hline 
		$0.8<z<1.4$ & $3086\pm286$ & $330\pm15$ & $-16\pm17$& $19.102154$ & $22$   \\ 
		\hline 
	\end{tabular}
\end{table}

\section{conclusion}

In this paper, we have investigated the properties of bulk flow across various redshift ranges in the context of $f(R)$ gravity, perturbed $f(R)$ gravity, and perturbed $f(R)$ gravity coupled with neutrinos. Our analysis reveals significant insights into the dynamics of large-scale cosmic flows and their alignment with prominent cosmic structures such as the Sloan Great Wall (SGW) and Ghidorah Supercluster (KGSc).

We observe that the inclusion of neutrinos in the perturbed $f(R)$ gravity model leads to a notable increase in the bulk flow velocity at all redshift ranges. This enhancement is particularly prominent in higher redshift ranges, with velocities exceeding $3000 \, \mathrm{km/s}$ in the $0.8 < z < 1.4$ range. Furthermore, the direction of the bulk flow in this model shows a striking alignment with the dark energy dipole, particularly at redshifts $z > 0.4$, where the bulk flow direction coincides almost exactly with the direction of the cosmic superclusters. In the redshift range $0.8 < z < 1.4$, for instance, the bulk flow direction aligns precisely with $(l, b) = (330^{\circ} \pm 15^{\circ}, -16^{\circ} \pm 17^{\circ})$, which is consistent with the direction of the dark energy dipole. This result implies a strong connection between the interaction of neutrinos and the underlying structure of the universe, potentially shedding light on the role of neutrinos in influencing cosmic acceleration.

At lower redshifts, such as $0.1 < z < 0.2$, the bulk flow is aligned with the direction of the Sloan Great Wall (SGW), while in the $0.4 < z < 0.6$ range, the flow aligns with the Ghidorah Supercluster (KGSc). This correlation between the bulk flow direction and major cosmic structures suggests that large-scale anisotropies are driven by gravitational forces associated with massive cosmic structures. The addition of neutrinos in the perturbed $f(R)$ gravity model further sharpens this alignment, providing a more accurate reflection of observed large-scale flows in the universe.

For redshifts below $z < 0.1$, we observe smaller bulk flow velocities, but the flow direction remains consistent with nearby cosmic structures. In the redshift range $0.001 < z < 0.016$, for instance, the bulk flow is influenced by structures like the Local Supercluster, with velocities typically lower than at higher redshifts. Even in this low redshift range, the introduction of neutrinos in the perturbed $f(R)$ gravity model still leads to a subtle increase in bulk flow velocity, suggesting that neutrino interactions play a role in cosmic flows even at small scales. Additionally, the bulk flow direction for $z < 0.1$ in the neutrino-coupled model is aligned with the Local Supercluster and similar structures, reflecting the influence of nearby gravitational sources.

In conclusion, our results highlight the crucial role of neutrinos in shaping the velocity and direction of bulk flows in the universe across both low and high redshift ranges. The alignment of the bulk flow with the dark energy dipole and the enhanced velocity in the neutrino-coupled model provide strong evidence of the impact of neutrinos on cosmic dynamics. These findings not only deepen our understanding of cosmic flows but also offer new avenues for probing the interactions between dark energy, neutrinos, and modified gravity models in shaping the evolution of the universe. Future studies should further explore these connections, with the goal of uncovering the precise mechanisms through which neutrinos influence large-scale cosmic structures.
\bibliography{sample631}{}
\bibliographystyle{aasjournal}



\section*{Appendix}

In this section, We investigate about the effect of the bulk flow on CMB poer spectrum.
Figure 11 illustrates the CMB power spectrum (\( C_\ell \)) as a function of the multipole moment (\( \ell \)) for various cosmological models, including \(\Lambda\)CDM, \(f(R)\) gravity, perturbed \(f(R)\), and perturbed \(f(R)\) with  neutrinos. The power spectrum reflects the distribution of temperature fluctuations in the CMB, and its shape is sensitive to both the underlying gravitational model and the presence of massive neutrinos. Here, we analyze the influence of these models on the CMB spectrum and their relation to the bulk flow velocity.

\subsection*{Effects of Models on the CMB Power Spectrum}
\begin{itemize}
	\item \textbf{\(\Lambda\)CDM Model:} As the baseline cosmological model, \(\Lambda\)CDM provides the standard predictions for the CMB spectrum, with well-defined peaks corresponding to baryon-photon acoustic oscillations. The low-\(\ell\) power is primarily determined by the Integrated Sachs-Wolfe (ISW) effect.
	
	\item \textbf{\(f(R)\) Gravity:} The inclusion of \(f(R)\) modifications alters the gravitational potential evolution, leading to slight enhancements in the ISW effect and a marginal shift in the peak heights. This results in increased low-\(\ell\) power and minor deviations in the higher multipoles.
	
	\item \textbf{Perturbed \(f(R)\):} Introducing perturbations in \(f(R)\) further amplifies the density fluctuations, which increases the relative heights of the acoustic peaks. This model also modifies the Silk damping scale, affecting the small-scale power (\(\ell > 500\)).
	
	\item \textbf{Perturbed \(f(R)\) + Neutrinos:} The coupling of massive neutrinos with perturbed \(f(R)\) gravity introduces significant changes. Massive neutrinos suppress small-scale structure growth due to their free-streaming behavior, which leads to damping at high multipoles. However, their gravitational interactions enhance the large-scale ISW effect, leading to the most pronounced increase in power at low-\(\ell\).
\end{itemize}

\subsection*{Bulk Flow Velocity and its Relation to CMB Power Spectrum}
The bulk flow velocity is directly related to the large-scale motion of matter in the universe and is imprinted in the low-\(\ell\) part of the CMB power spectrum. In the perturbed \(f(R)\) + neutrinos model, the bulk flow velocity is the highest among all models. This can be attributed to the following factors:
\begin{itemize}
	\item \textbf{Enhanced ISW Effect:} The gravitational potentials in the perturbed \(f(R)\) model evolve dynamically, and the presence of massive neutrinos further delays the decay of these potentials. This prolongs the ISW effect and increases the low-\(\ell\) power, which is directly linked to bulk flow velocity.
	\item \textbf{Neutrino Gravitational Coupling:} Massive neutrinos contribute to the gravitational field and interact with perturbations in \(f(R)\) gravity. This interaction enhances the large-scale flows, leading to higher bulk flow velocities compared to other models.
\end{itemize}

\subsection*{Conclusion}
The perturbed \(f(R)\) + neutrinos model exhibits the highest bulk flow velocity due to the synergistic effects of dynamic gravitational potentials and massive neutrino interactions. These results highlight the sensitivity of the CMB power spectrum to modifications in gravity and the properties of neutrinos, emphasizing the importance of low-\(\ell\) power in probing large-scale cosmic flows.

\begin{figure}
	\centering
	\includegraphics[scale=0.58]{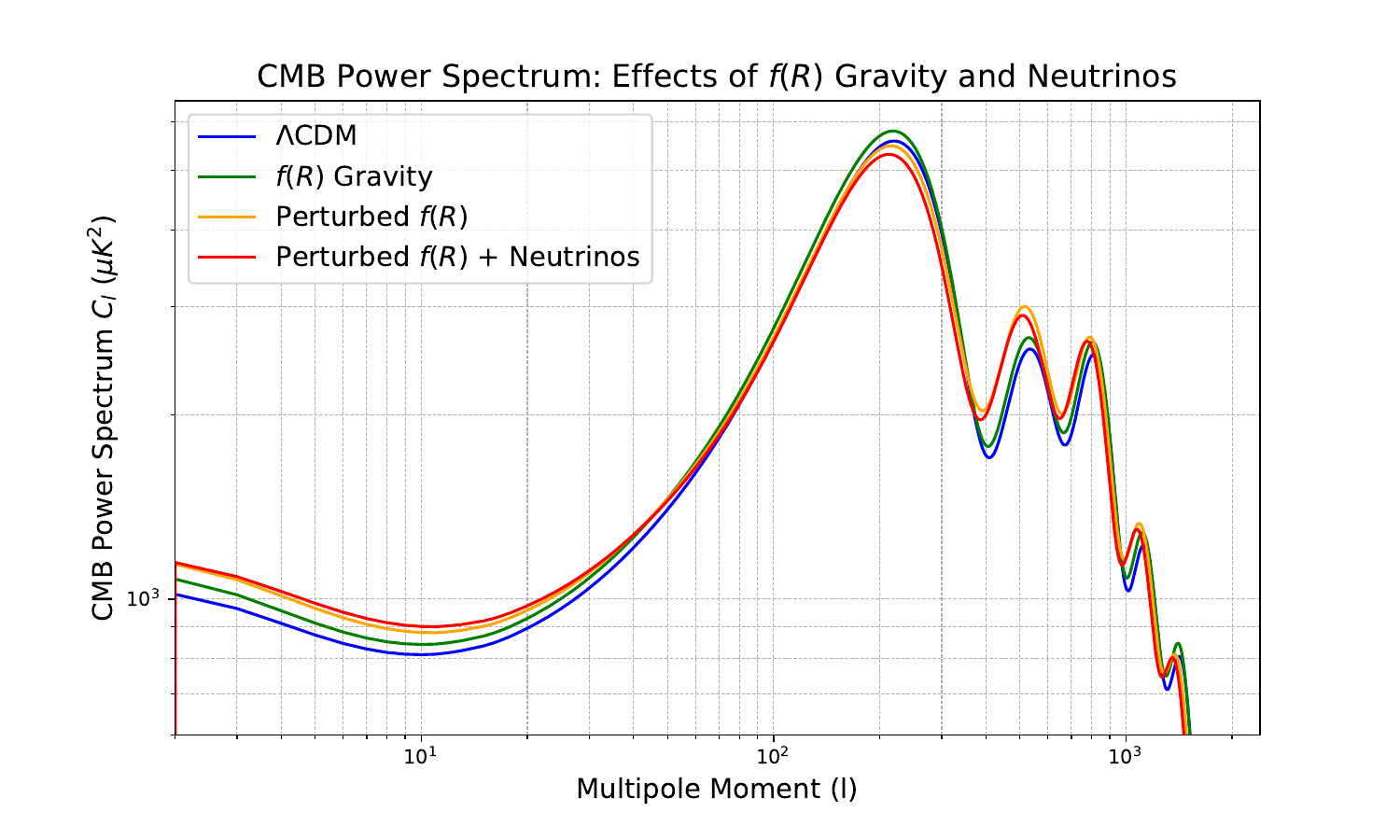}
	\caption{Comparison of different models on the CMB power spectrum.}
	\label{fig:CMB_comparison}
\end{figure}

\end{document}